\documentclass[conference]{IEEEtran}
\pdfoutput=1
\usepackage{cite}
\usepackage{amsmath,amssymb,amsfonts}
\usepackage{algorithmic}
\usepackage{graphicx}
\usepackage{textcomp}
\usepackage{xcolor}
\usepackage[hyphens]{url}

\usepackage{soul}
\usepackage{epstopdf}
\usepackage{fancyhdr}
\usepackage{color}
\usepackage{soul}
\usepackage{subfig}
\usepackage{epsfig}
\usepackage{alltt}
\usepackage{enumitem}
\usepackage[normalem]{ulem}
\usepackage{float}
\usepackage[]{algorithm2e}
\usepackage{lipsum}
\usepackage{calc}
\usepackage{multirow}
\usepackage{pifont}
\usepackage{adjustbox}
\usepackage{booktabs}
\usepackage{listings}
\usepackage{tikz}
\usepackage{diagbox}
\usepackage{tabularx}
\usepackage[framemethod=tikz]{mdframed}
\usepackage{colortbl}
\usepackage{formatting/macros}
\usepackage{formatting/results}
\usepackage{makecell}
\usepackage{subfig}
\usepackage{epsfig}
\usepackage{tcolorbox}
\usepackage[numbers,sort&compress]{natbib}

\usepackage{balance}
\usepackage{amsmath}
\usepackage{amssymb}
\usepackage[hidelinks,colorlinks=true,linkcolor=blue,citecolor=black]{hyperref}

\def\BibTeX{{\rm B\kern-.05em{\sc i\kern-.025em b}\kern-.08em
    T\kern-.1667em\lower.7ex\hbox{E}\kern-.125emX}}

\begin{document} 
\title{Heterogeneous Acceleration Pipeline for Recommendation System Training}

\author{
	\IEEEauthorblockN{Muhammad Adnan$^\dagger$ \quad Yassaman Ebrahimzadeh Maboud$^\dagger$ \quad Divya Mahajan$^\star$ \quad Prashant J. Nair$^\dagger$}\\
 
	\IEEEauthorblockA{
		$^\dagger$The University of British Columbia \quad \quad $^\star$Georgia Institute of Technology} \\

    \IEEEauthorblockA{\{adnan, yassaman, prashantnair\}@ece.ubc.ca \quad \quad divya.mahajan@gatech.edu}
}


\maketitle

\begin{abstract}
Recommendation models rely on deep learning networks and large embedding tables, resulting in computationally and memory-intensive processes. These models are typically trained using hybrid CPU-GPU or GPU-only configurations. The hybrid mode combines the GPU's neural network acceleration with the CPUs' memory storage and supply for embedding tables but may incur significant CPU-to-GPU transfer time. In contrast, the GPU-only mode utilizes High Bandwidth Memory (HBM) across multiple GPUs for storing embedding tables. However, this approach is expensive and presents scaling concerns.

This paper introduces \hotline, a heterogeneous acceleration pipeline that addresses these concerns. \hotline develops a data-aware and model-aware scheduling pipeline by leveraging the insight that only a few embedding entries are frequently accessed (popular). This approach utilizes CPU main memory for non-popular embeddings and GPUs' HBM for popular embeddings. To achieve this, \hotline accelerator fragments a mini-batch into popular and non-popular micro-batches ($\mu$-batches). It gathers the necessary working parameters for non-popular $\mu$-batches from the CPU, while GPUs execute popular $\mu$-batches. The hardware accelerator dynamically coordinates the execution of popular embeddings on GPUs and non-popular embeddings from the CPU's main memory. Real-world datasets and models confirm \hotline's effectiveness, reducing average end-to-end training time by \avgperfimprfourgpudlrm compared to Intel-optimized CPU-GPU DLRM baseline.
\end{abstract}


\section{Introduction}
\label{sec:introduction}

Recommendation models constitute a crucial and widely deployed class of machine learning (ML) workloads~\cite{acun2020understanding}. These models employ compute-intensive neural networks and memory-intensive embedding tables to store user and item features~\cite{facebookemb}. With the increasing number of interactions between users and items, the size of these tables is expected to grow significantly. Production-scale models can already reach several terabytes and contain trillions of parameters~\cite{dsi, zhao2020distributed, dlrm2022}.

\begin{figure}
  \centering
    \subfloat[Hybrid CPU-GPU mode]{
	\begin{minipage}[t]{1\linewidth}
	   \centering
	   \includegraphics[width=\textwidth]{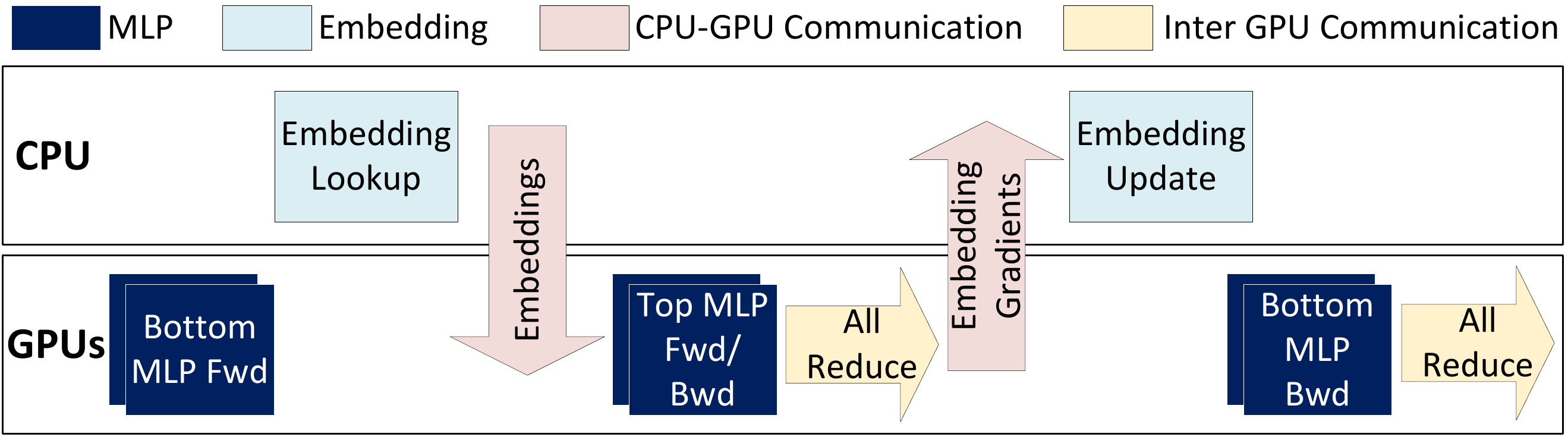}
	   \vspace{-0.15in}
	   \label{fig:baselineflowhybrid}
	\end{minipage}}
	\\
  \subfloat[GPU-only mode]{
	\begin{minipage}[t]{1\linewidth}
	   \centering
	   \includegraphics[width=\textwidth]{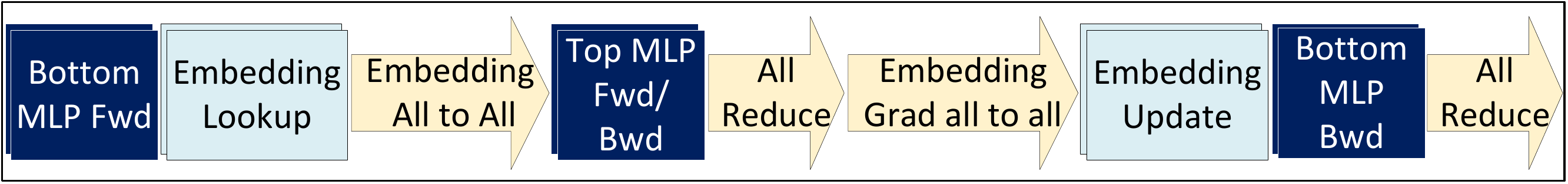}
	   \vspace{-0.15in}
	   \label{fig:baselineflowgpu}
	\end{minipage}}
\caption{The execution flow of a typical recommendation model in the hybrid CPU-GPU and GPU-only. Due to their large sizes, the embedding tables are stored and processed on CPUs. The GPUs process the neural layers.}
\label{fig:baselineflow}
\vspace{-0.2in}
\end{figure}

The deep Learning Recommendation Model (DLRM) and the Time-Based Sequence Model (TBSM) are popular commercial models. These models are typically trained using either a hybrid CPU-GPU mode or a GPU-only mode~\cite{dlrm, tbsm}. In the hybrid mode (Figure~\ref{fig:baselineflowhybrid}), the CPU provides memory capacity for the embedding entries, while GPUs offer high-throughput data-parallel neural network execution~\cite{largescalegputraining}. However, this mode suffers from inefficiencies due to three reasons: (1) GPUs rely on the CPU to provide all embeddings, (2) the low-bandwidth CPU memory acts as a bottleneck, and (3) the CPU's execution of the embedding logic prevents full GPU utilization throughout the training process.

Alternatively, the GPU-only mode, illustrated in Figure~\ref{fig:baselineflowgpu}, employs multiple GPUs to store a single copy of the embeddings and trains in a model-parallel manner~\cite{hugectr}. However, this approach necessitates continuous all-to-all communication between GPUs to share their embeddings. Additionally, in this mode, one would need to grow the GPUs to enable the training of larger datasets. For instance, the Terabyte dataset requires at least \emph{four} NVIDIA V100 GPUs to fit its embeddings. Ideally, even larger applications must be enabled using fewer GPUs.

To overcome these limitations of existing training modes, this paper introduces a novel heterogeneous acceleration pipeline called \hotline. The primary goal of \hotline is to fully exploit GPUs' compute throughput and CPU-based memory capacity without encountering any communication or computation bottlenecks. By combining the advantages of both the hybrid and GPU-only modes, \hotline utilizes the GPU-only mode for the entire training process. It leverages the CPU-based main memory to store the majority of embeddings.

This is achieved through an innovative hardware accelerator that pipelines the embedding gathering operation of the CPU together with the compute on GPUs. This approach ensures that GPUs are continuously fed from either the CPU-based memory subsystem or their High Bandwidth Memories (HBM), avoiding stalls in the process. As a result, \hotline provides a scalable throughput-optimized solution. Broadly, the \hotline acceleration pipeline relies on two key insights.

\niparagraph{1. Access-aware Embedding Layout in Memory:} \hotline leverages the observation that real-world recommender systems exhibit a high skew in popularity, causing certain embedding entries to be accessed significantly more frequently than others \cite{fae,recshard,wave,popsimilar,bagpipe}. These frequently accessed embeddings, referred to as \emph{\highaccess entries}, have a small memory footprint but are computationally significant. To take advantage of this access property, \hotline proposes an optimized \emph{access-aware memory layout for embeddings}. The hardware accelerator in \hotline dynamically classifies \highaccess embeddings and places them on the GPU memory, while not-\highaccess embeddings are stored in the CPU main memory. The hardware accelerator periodically monitors the access pattern to ensure it captures the most up-to-date trends in the training data.

\niparagraph{2. Layout-aware Runtime Scheduling:} \hotline employs a dynamic runtime scheduler to achieve optimal compute throughput with the new memory placement. This scheduler divides a mini-batch into two micro-batches ($\mu$-batches), and subsequently, these $\mu$-batches are categorized into two groups. The first category comprises popular inputs that exclusively access \highaccess embeddings and are directly scheduled onto the GPUs. Remarkably, we find that approximately 75\% of the inputs fall into the first category across a wide range of models. On the other hand, the second category consists of the remaining inputs that may access both \highaccess and not-\highaccess embeddings. If an input accesses even a single non-\highaccess embedding, it is classified as a non-popular input. For the first category of popular inputs, all the required embeddings for the micro-batch are directly scheduled onto the GPUs. In contrast, \hotline gathers the not-\highaccess embeddings from the CPU memory for the second category. However, accessing the CPU's main memory can stall the pipeline in such a system. This is because the not-\highaccess embeddings are stored in the CPU's main memory and are in the critical path.

To overcome this challenge, \hotline introduces a novel hardware accelerator that schedules the $\mu$-batches in a data-aware and model-aware manner. By doing so, \hotline ensures that GPUs are not starved, allowing the system to gather the not-\highaccess embeddings from the CPU memory while executing the \highaccess inputs on the GPUs.

\vspace{0.05in}
\noindent \textbf{Contributions:} This work makes three key contributions:
\begin{enumerate}[noitemsep]
\item Identifies \highaccess embeddings \emph{dynamically} at runtime with negligible overhead.
\item Offers a dynamic data and model-aware scheduler to efficiently pipeline the mini-batch dispatch onto GPUs while concurrently obtaining not-\highaccess embeddings from the main memory.
\item Offers a runtime framework that increases training throughput by stitching the training process of the recommender model across CPUs and GPUs.
\end{enumerate}

We evaluated \hotline with publicly available deep learning (DLRM) and time-sequence (TBSM) based recommendation models and compared our approach against two baselines - hybrid CPU-GPU and GPU-only baseline.

We compared Hotline against state-of-the-art deep learning frameworks such as XDL~\cite{xdl}, FAE~\cite{fae}, and Intel-optimized DLRM~\cite{dlrm-intel}. On average, \hotline provides \avgperfimprfourgpuxdl speedup over the 4-GPU XDL, \avgperfimprfourgpufae over FAE, and \avgperfimprfourgpudlrm speedup over Intel optimized DLRM. It is noteworthy that \hotline only rearranges inputs in a single mini-batch, but the updates to the model are performed at parity with the baseline. Thus, \hotline does not impact the accuracy or training fidelity of the model. \hotline could train larger models, such as Criteo Terabyte, with a single GPU, whereas the GPU-only baseline required at least 4 GPUs to store its embeddings.
\section{Recommendation Systems}
\label{sec:background}

\begin{figure}[b!]
        \vspace{-0.1in}
	\centering
	\includegraphics[width=0.3\textwidth]{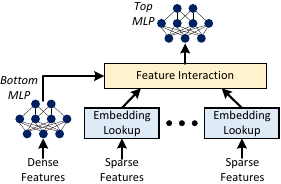}
	\caption{General structure of a deep-learning-based recommendation model~\cite{dlrm,neuralcf,tbsm}. It consists of compute-bound neural networks and memory-bound embedding tables.}
	\label{fig:recomodel}
\end{figure}

Figure~\ref{fig:recomodel} illustrates the general structure of deep-learning-based recommendation models, which rely on two types of inputs: dense and sparse. Dense inputs are continuous features, such as the user's age, while sparse inputs represent categorical features, such as the user's location or videos they have liked. The neural network component processes dense inputs using Multi-Layer Perceptron (MLP) techniques, while massive embedding tables handle sparse inputs. Each embedding table represents a categorical feature, with the number of rows corresponding to the possible items associated with that feature. An MLP processes the outputs of both the dense and sparse inputs to generate a prediction, such as the likelihood of clicking or click-through rate (CTR). 

\subsection{Training Setup}
\label{subsec:training_setup}

Large recommender models can be trained using two distributed modes: the hybrid mode and the data-parallel mode. In the hybrid mode, embeddings are stored and gathered on the CPU, while neural networks are executed on GPUs in a data-parallel manner. However, the training throughput of the hybrid mode is often limited due to substantial data transfers and reliance on low-bandwidth CPU main memory.

\begin{figure}[t!]
	\centering
	\includegraphics[width=0.85\columnwidth]{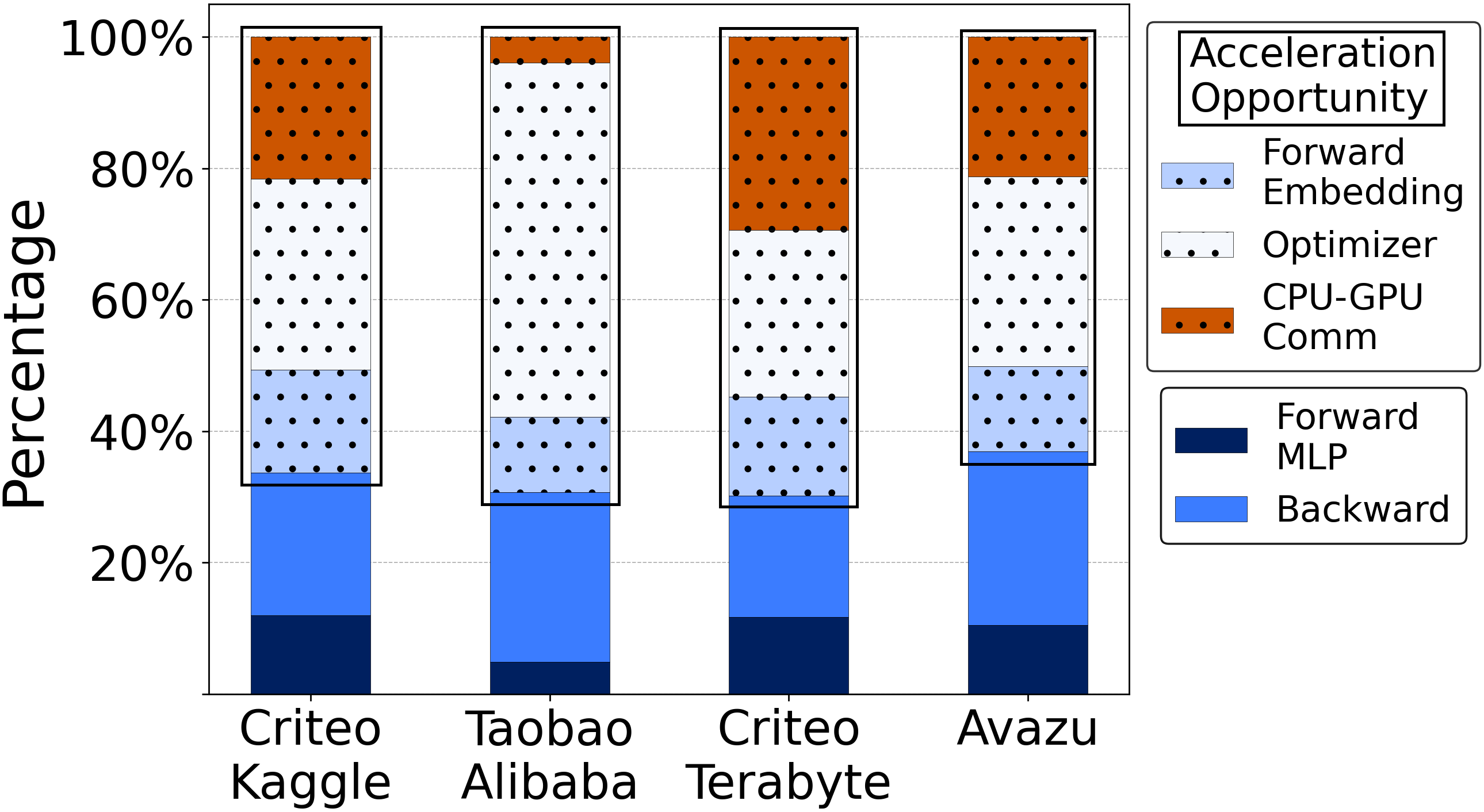}
	\caption{The breakdown of the training time for an Intel-optimized DLRM with 4-GPU in a hybrid CPU-GPU training setup. The dotted parts of the bar are executed on the CPU and present an opportunity for GPU-based acceleration.}
	\label{fig:motivation}
    \vspace{-0.2in}
\end{figure}

\subsubsection{Hybrid CPU-GPU Mode}
\label{subsubsec:hybrid_cpu_gpu}

Figure~\ref{fig:motivation} illustrates the distribution of training time across four real-world models and datasets. The results highlight that embedding operations, such as embedding-lookup in the forward pass, updating embeddings in the optimizer, and CPU-GPU communication, can account for up to 75\% of the training time in large datasets, like Criteo Terabyte.

\subsubsection{Single Node GPU-only Mode}
\label{subsubsec:single_gpu_only}

\begin{figure}[b!]
	\vspace{-0.05in}
	\centering
	\includegraphics[width=0.9\columnwidth]{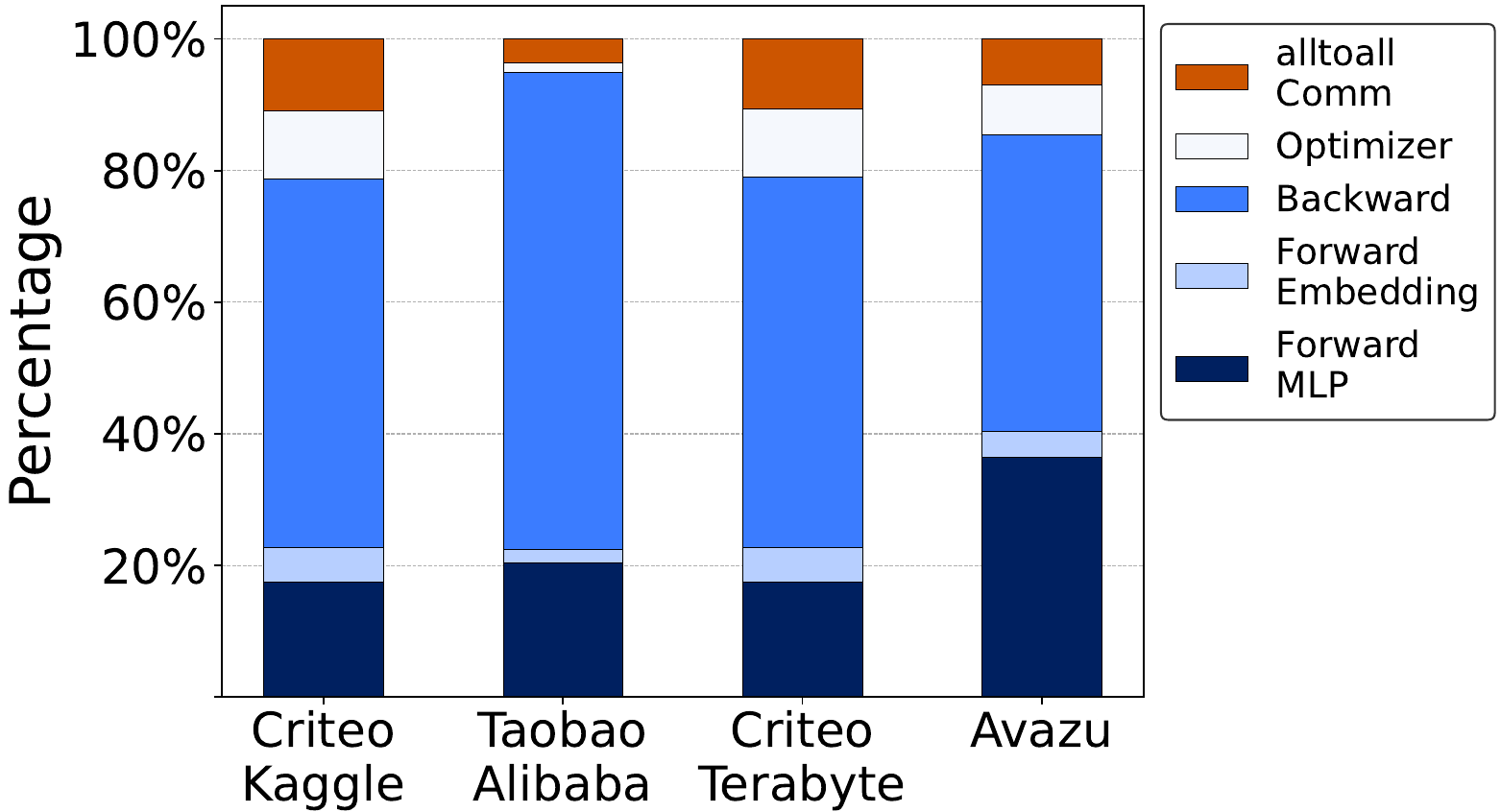}
	\caption{The breakdown of the training time for DLRM in single node GPU-only training setup. The single-node setup uses NVLink interconnect across four GPUs.}
	\label{fig:motivation_gpu_only}
\end{figure}

In the GPU-only mode, multiple GPUs are used to store all embeddings and perform data-parallel neural network execution. However, this mode experiences low compute utilization primarily because recommendation models grow with the size of the embedding tables, resulting in a larger memory footprint rather than an increase in neural compute~\cite{dsi, zhao2020distributed, dlrm2022}. Consequently, the GPU devices must scale with the size of the embedding table rather than the neural network's compute intensity. Figure~\ref{fig:motivation_gpu_only} illustrates the breakdown of training time for four real-world datasets using a single node GPU-only system with NVLink~\cite{nvlink} interconnect.

In a single node GPU-only system, transferring embeddings across all devices requires {\fontfamily{qcr}\selectfont all-to-all} collectives. For instance, in a 4-GPU system, we observed that this step consumes nearly 12\% of the total training time even after employing the fast NVLink interconnect. As the number of nodes increases, the communication time also grows, potentially limiting scalability and becoming the training bottleneck~\cite{zionex-fb}.

\begin{figure}[t]
	\centering
	\includegraphics[width=\columnwidth]{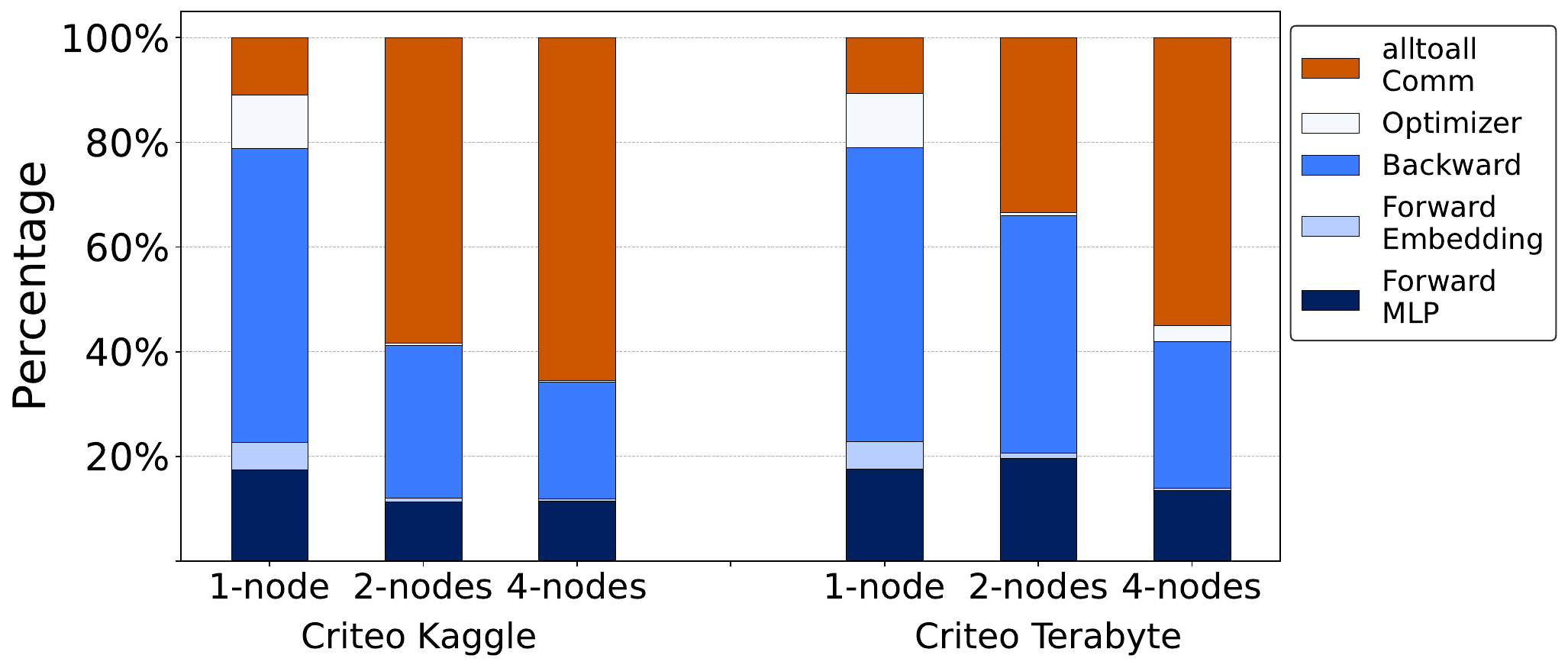}
	\caption{The training time breakdown for DLRM in a multi-node GPU-only setup with four GPUs per node. The GPUs use NVLink for intra-node GPU connections. The multi-node setup uses 100Gbps InfiniBand for inter-node connectivity.}
	\label{fig:multi_node_gpu_only}
\end{figure}

\subsubsection{Multi Node GPU-only Mode}
\label{subsubsec:multi_gpu_only}

Distributed training across multiple nodes exacerbates communication time, particularly with {\fontfamily{qcr}\selectfont all-to-all} collectives. In our Dell EMC C4140 nodes, InfiniBand links provide a bandwidth of only 100Gbps, while NVLink offers 2400Gbps for Nvidia-V100 GPUs. This disparity makes communication a major bottleneck. As shown in Figure~\ref{fig:multi_node_gpu_only}, communication costs now exceed 50\% of the multi-node training time.

\subsection{Popularity in Training Inputs}

\begin{figure}[b!]
	\centering
	\begin{minipage}[t]{0.45\textwidth}
	   \centering
	   \includegraphics[width=\textwidth]{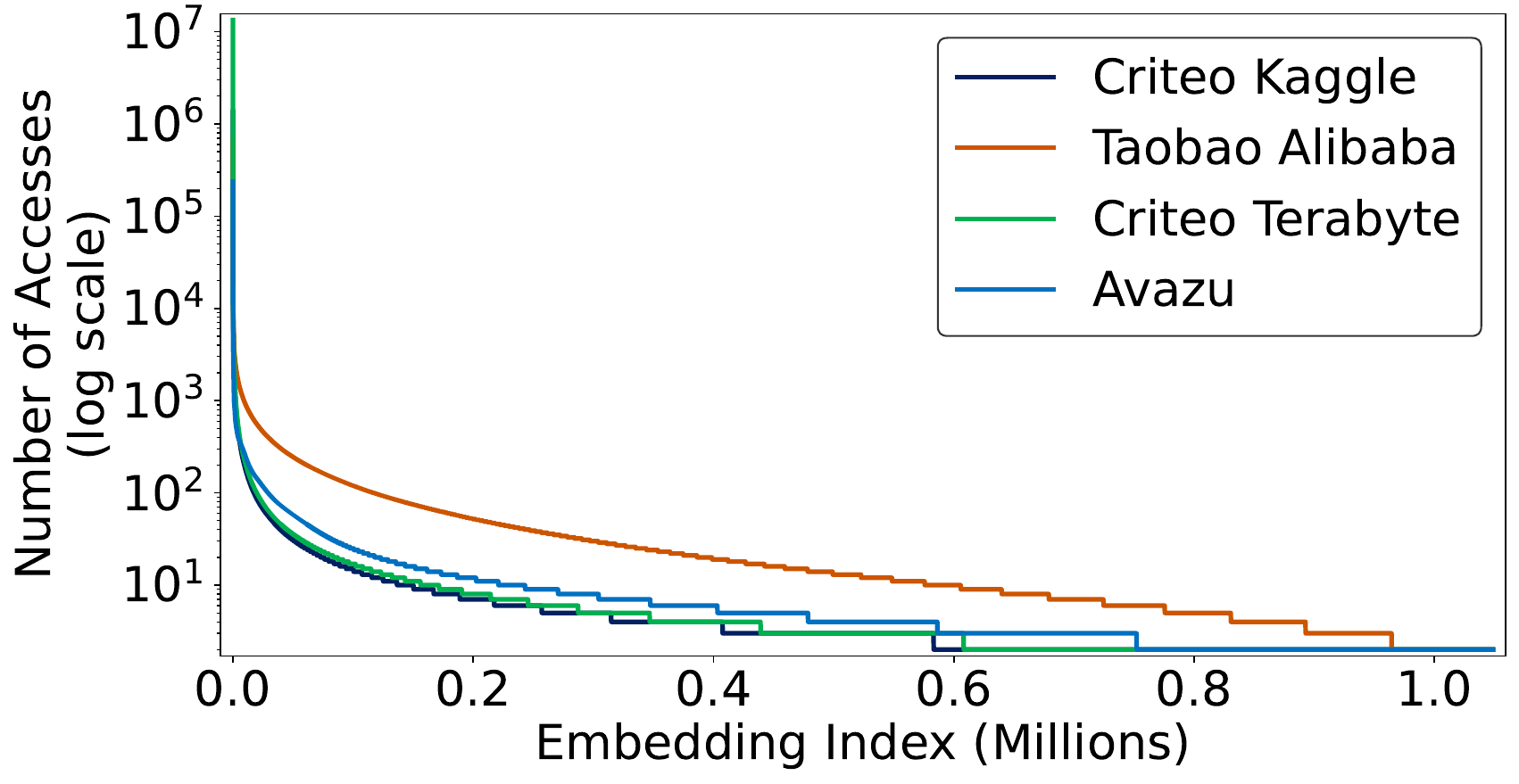}
	\end{minipage}
    \\
    \begin{minipage}[t]{0.45\textwidth}
	   \centering
	   \includegraphics[width=0.9\textwidth]{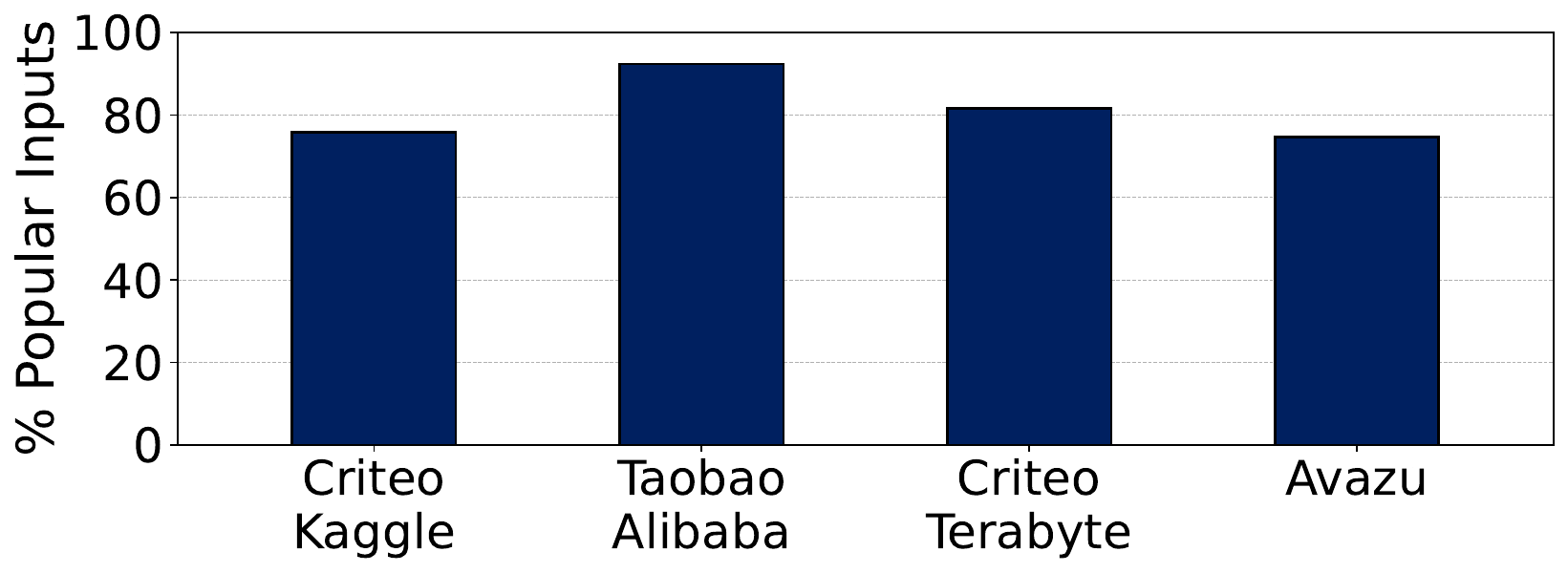}
	\end{minipage}
	\caption{Number of accesses per embedding entry per one training epoch. Inputs that account for at least 1-in-every-100000 embedding accesses are labelled as \emph{popular}.}
	\label{fig:ip_emb_acc}
\end{figure}

The \hotline framework leverages a fundamental characteristic of recommendation models, where specific users and items exhibit significantly higher popularity than others. This phenomenon leads to certain embeddings being accessed far more frequently than others, as illustrated in Figure~\ref{fig:ip_emb_acc} across various real-world datasets. Typically, a small number of \highaccess embeddings can receive over 100$\times$ more access than others, catering to over 75\% of inputs with only approximately 512~MB of embeddings. Nevertheless, adapting to changes in input popularity poses a complex challenge.

\section{Challenges and Insights}

The \highaccess embeddings have a small memory footprint and serve the majority of inputs, making it feasible to place these \highaccess embeddings locally across GPUs. This approach could eliminate the need to involve CPUs for most inputs, leading to significant performance benefits. However, this approach faces three key challenges.

\niparagraph{Challenge 1 -- Embeddings in CPUs and GPUs:} 

During the training process, mini-batches of input data access a mix of \highaccess and non-\highaccess embeddings, which are stored across both the CPU main memory and the HBM of GPUs. Consequently, some embeddings from the CPU's main memory must be collected and transmitted to the GPU for embedding computations.

\finding{\textbf{Our Approach}: \hotline partitions each mini-batch into two micro-batches ($\mu$-batches). The inputs in a $\mu$-batch either access \emph{only} \highaccess embeddings or any arbitrary embeddings. First, \hotline schedules the $\mu$-batches that access \emph{only} \highaccess embeddings on the GPU(s) for execution. Concurrently, it collects the parameters for the $\mu$-batches that access embeddings from the CPU memory.}

\niparagraph{Challenge 2 -- Segregation and Scheduling with CPU:}
\label{sec:segregation_and_scheduling}
Achieving efficient mini-batch segregation and parameter gathering can be accomplished using CPUs and GPUs instead of hardware accelerators. However, GPUs are not optimized for fine-grained mini-batch segregation. To address this, CPU-based multi-processing can be employed for mini-batch segregation, parameter gathering, and scheduling.
\begin{figure}[h!]
	\centering
	\includegraphics[width=0.95\columnwidth]{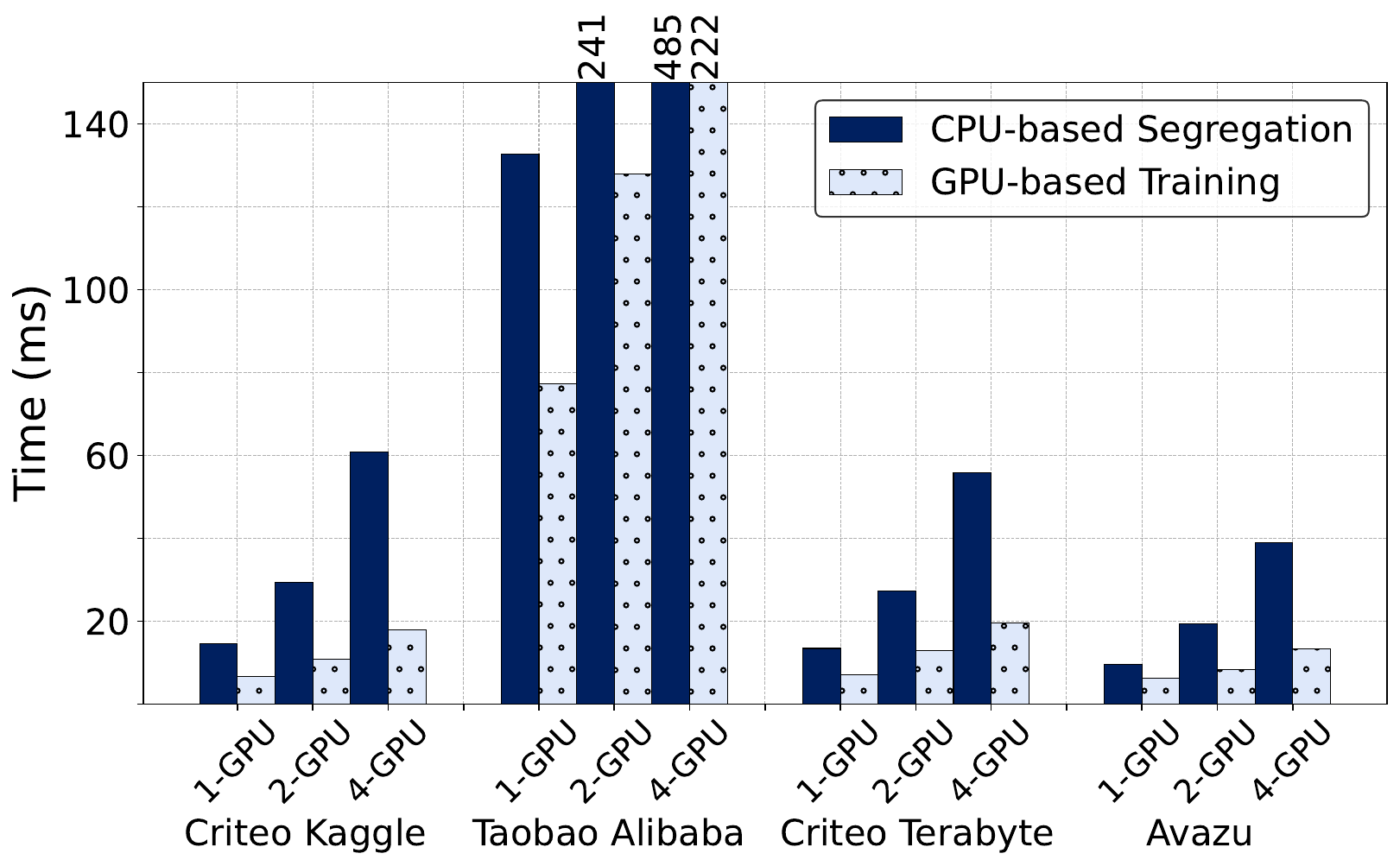}
	\caption{CPU segregation and scheduling time for a mini-batch using Intel Xeon CPU while the V100 GPU(s) training on a mini-batch. We use mini-batches of 1K, 2K, and 4K inputs for 1, 2, and 4-GPU execution, respectively. Each mini-batch contains two $\mu$-batches (popular and non-popular).}
	\label{fig:hotline_cpu_algo}
\end{figure}

Our study, as illustrated in Figure~\ref{fig:hotline_cpu_algo}, revealed that even when utilizing all CPU cores, an Intel Xeon CPU exhibits a mini-batch segregation latency up to 2.5$\times$ higher than that of NVIDIA-V100 GPU(s) single mini-batch GPU-based training. This is because CPU-based segregation necessitates numerous memory look-ups to determine if an input is high access.

\begin{figure}[t!]
	\centering
	\includegraphics[width=0.85\columnwidth]{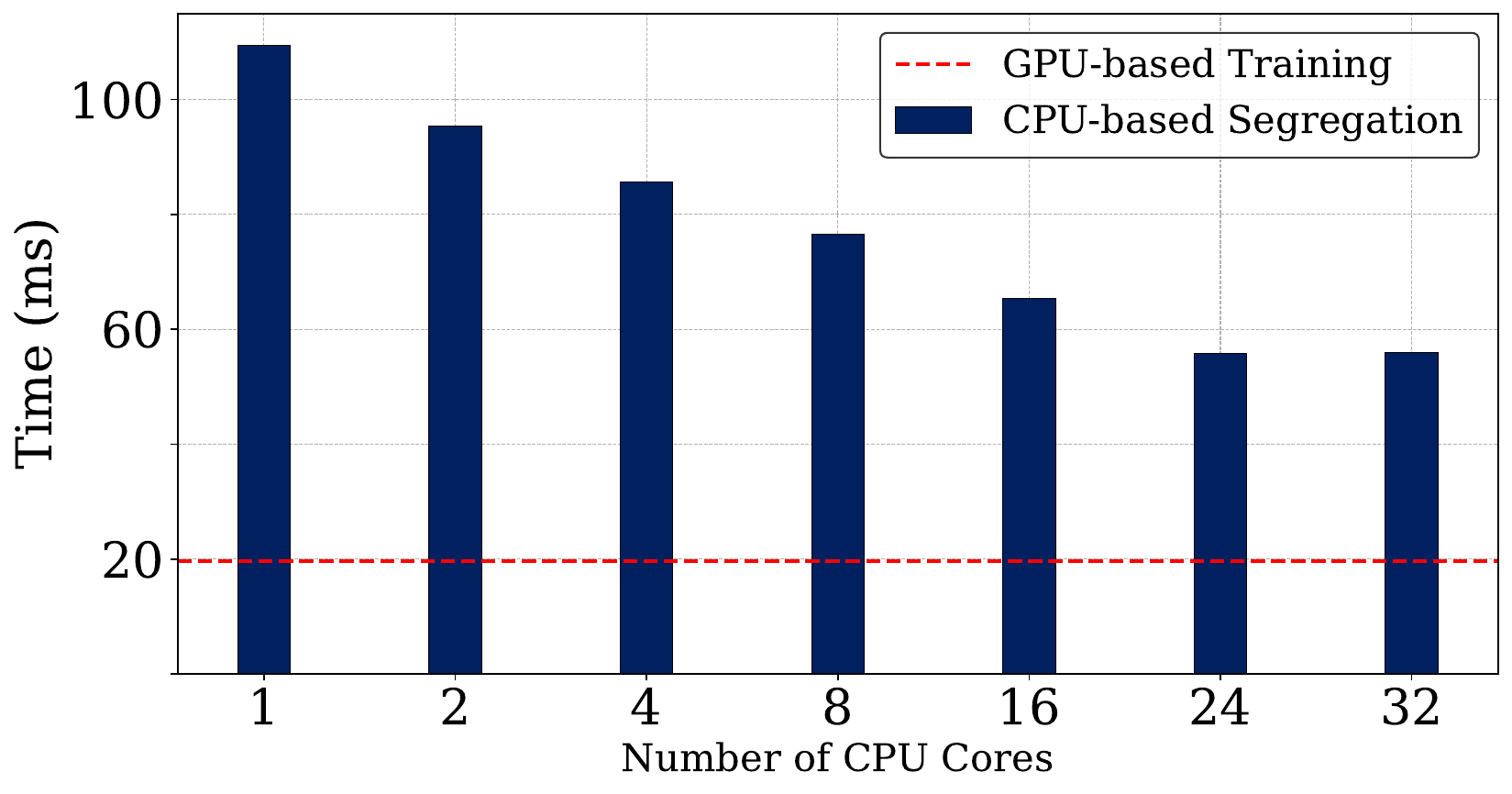}
	\caption{The wall-clock time by varying CPU cores for segregating a 4K input mini-batch of the Criteo Terabyte dataset. The time overhead of CPU-based segregation plateaus with an increasing number of cores.}
	\label{fig:multicore_cpu_algo}
\end{figure}

We investigated the bottleneck in CPU-based segregation by varying the number of CPU cores for segregating a Criteo Terabyte dataset mini-batch. Figure~\ref{fig:multicore_cpu_algo} illustrates that adding cores initially decreases segregation time slightly, but beyond 24 cores, segregation time plateaus. This indicates that the issue lies with parallel memory accesses from the CPU cores rather than CPU compute throughput. Therefore, segregation is memory bound, and even hardware like the Data Streaming Accelerator (DSA) (within Intel Sapphire Rapids), designed for data copying and transformation, would not alleviate the issue due to its inability to handle parallel memory lookups.

As the mini-batch size increases, the processing overhead and latency for CPU-based segregation also increase proportionately. CPUs cannot actively segregate and schedule popular and non-popular $\mu$-batches \emph{before} the GPUs finish their execution. Consequently, our experiments demonstrate that GPUs remain idle for over 50\% of the training time.

\finding{\textbf{Our Approach}: \hotline introduces a novel accelerator that utilizes parallel lookup engines capable of performing fine-grained tasks. These tasks include determining whether an input is a high-access or low-access value, fragmenting the mini-batches to form new $\mu$-batches, and enabling efficient parameter gathering for the $\mu$-batches. With the help of this accelerator, the acceleration pipeline can segregate and schedule the non-popular $\mu$-batch as soon as the GPUs finish executing the popular $\mu$-batch.}

\begin{figure}[t!]
  \centering
	\begin{minipage}[t]{0.34\linewidth}
	   \centering
	   \includegraphics[width=\textwidth]{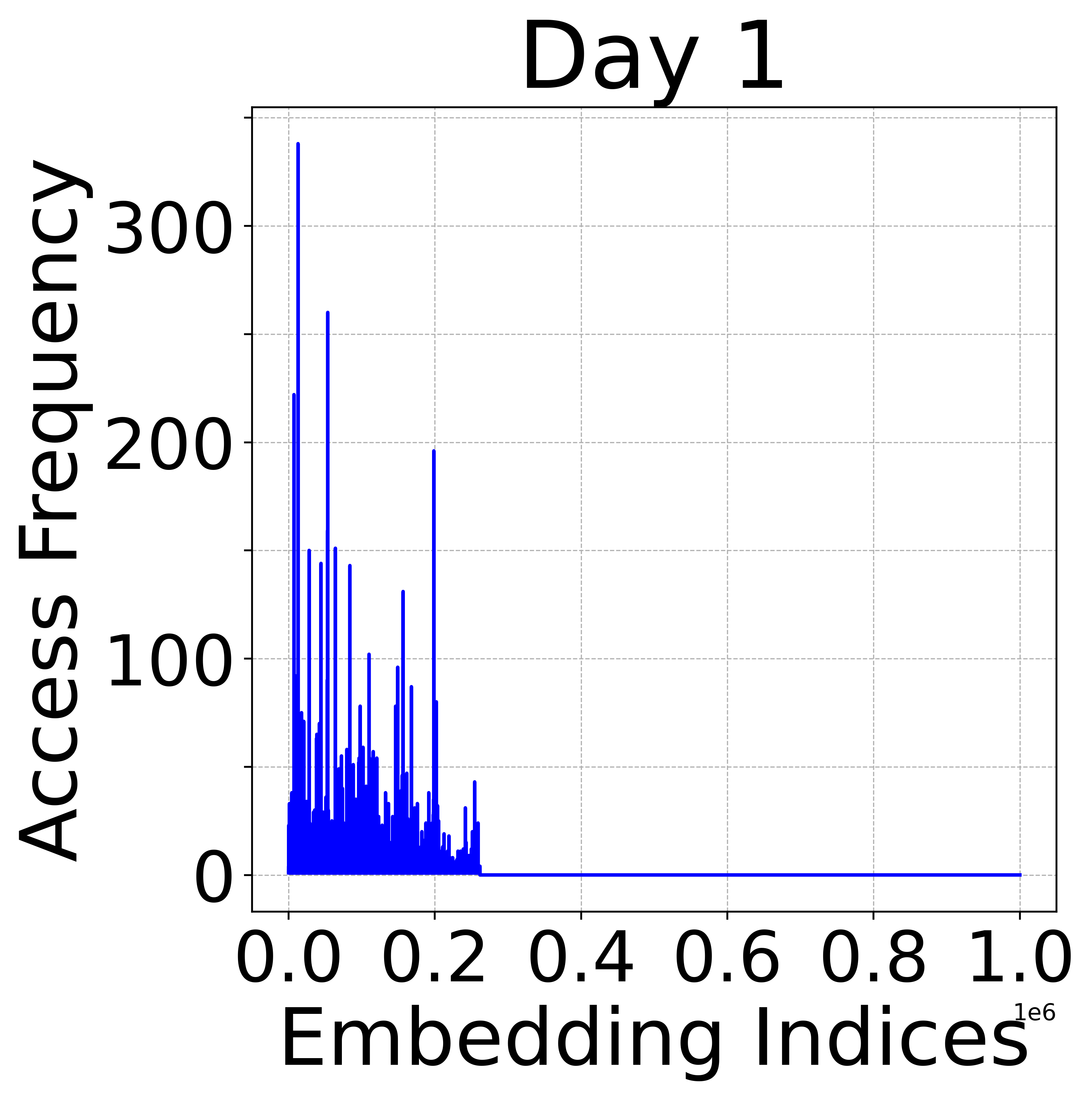}
	\end{minipage}
	\begin{minipage}[t]{0.3\linewidth}
	   \centering
	   \includegraphics[width=\textwidth]{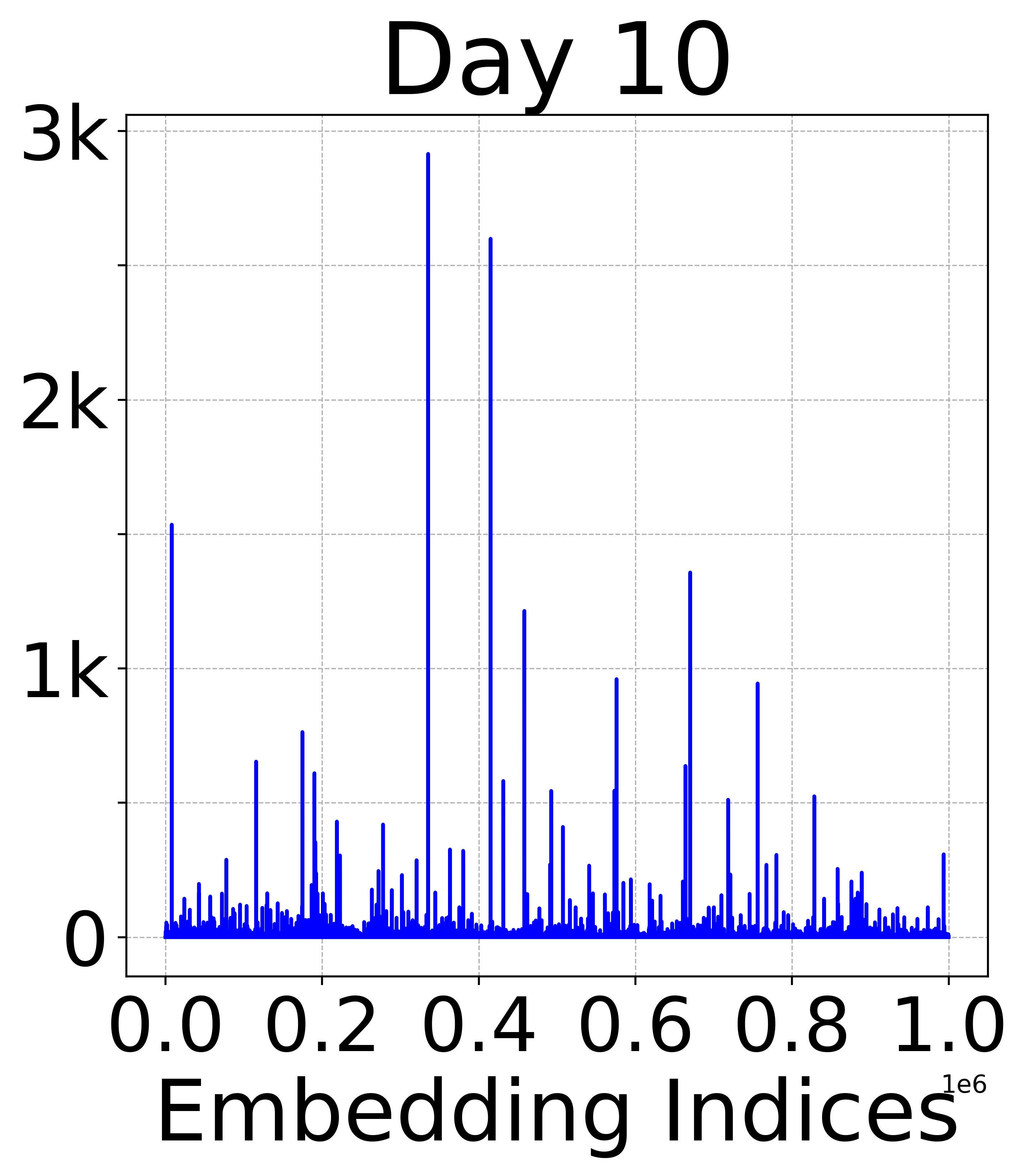}
	\end{minipage}
	\begin{minipage}[t]{0.3\linewidth}
	   \centering
	   \includegraphics[width=\textwidth]{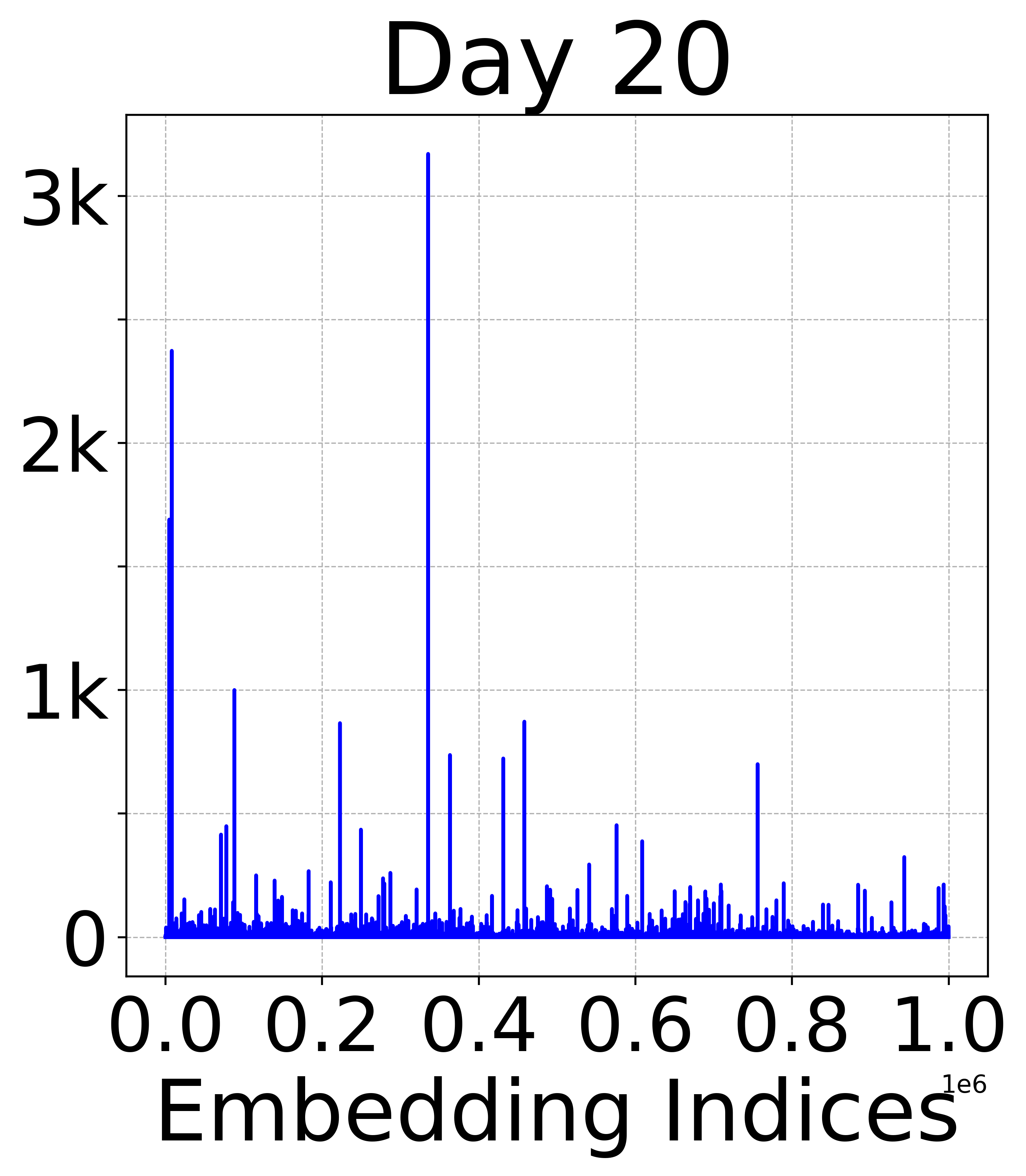}
	\end{minipage}
\caption{Evolving skew in training data across days for Terabyte dataset (Embedding Table 20). Thus, popular embeddings vary sometimes as frequently as a few hours.}
\label{fig:evolving_skew}
\end{figure}

\niparagraph{Challenge 3 -- Evolving Access Skews:}
\label{subsec:challenge3}
Prior studies have used an \emph{offline} profiler~\cite{fae, recshard} to identify \highaccess embeddings. Some of these studies do not account for this overhead, up to 15\%, in their training times~\cite{fae}. They also assume that the training data is available before training and that the set of \highaccess embeddings does not change over time. However, training data in the recommender models is mostly structured as user activity across some finite time. As user behavior changes rapidly every few hours or days, static profiling may not quickly identify the corresponding shift embedding accesses~\cite{bagpipe}. Figure~\ref{fig:evolving_skew} shows the change in user behavior for the Criteo Terabyte dataset across days.

\finding{\textbf{Our Approach}: To address the aforementioned issues, \hotline adopts a dynamic approach that \emph{samples} a small fraction of inputs (usually 5\%) to identify \highaccess embeddings. This minimizes the profiling overheads to be $\leqslant$5\% while enabling efficient tracking of \highaccess embeddings across mini-batches. Furthermore, the accelerator continuously re-calibrates the \highaccess embeddings to adapt to changes in training data.}
\section{The Hotline System}
\label{sec:system}

\begin{figure*}[h]
	\centering
	\includegraphics[width=0.9\textwidth]{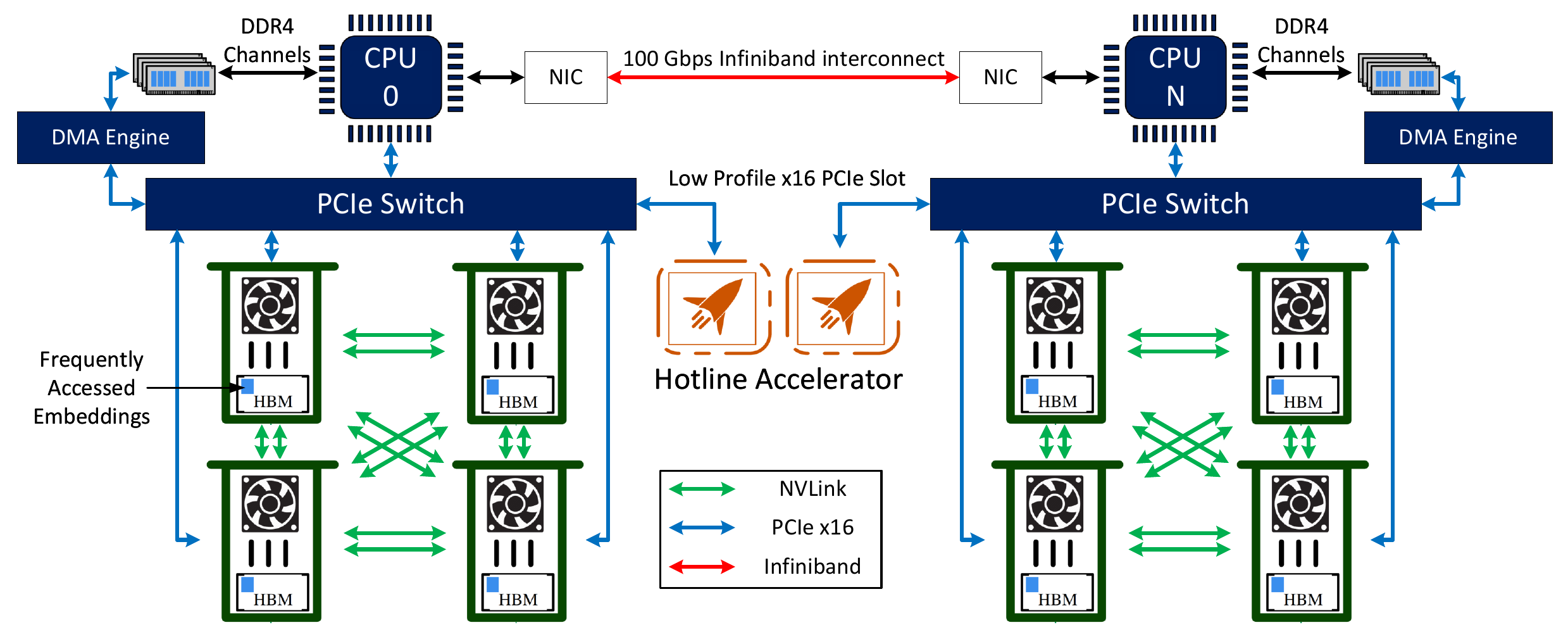}
	\caption{The \hotline system features an accelerator situated between the CPU and GPU(s), responsible for accessing the main memory to retrieve training inputs and embeddings, which are then efficiently relayed to the GPU(s). In multi-node distributed training, each node utilizes its own \hotline accelerator to oversee and execute parameter aggregation for its mini-batch.}
	\label{fig:hotlinemultinodesystem}
\end{figure*}

The system comprises an InfiniBand-based multi-core server. Each node has multiple GPU devices with inter-GPU communication achieved via NVLink~\cite{nvlink}. The \hotline system places the accelerator on the \emph{low profile PCIe slot} that GPUs do not use, enabling it to access the DMA engine through the PCIe switch and communicate directly with the CPU memory as shown in Figure~\ref{fig:hotlinemultinodesystem}. Notably, the \hotline system requires no modifications to the CPU and GPU devices. This system operates in two phases: 

\niparagraph{1. The Learning Phase:} The \hotline accelerator actively determines the \highaccess embeddings at runtime. To achieve this, the accelerator performs mini-batch sampling in the first epoch. Our experiments demonstrate that sampling just 5\% of the mini-batches is sufficient to identify over 90\% of the \highaccess embeddings. Based on the sampled mini-batches, the accelerator progressively classifies the accessed embeddings as either \highaccess or non-\highaccess. Subsequently, the contents of the \highaccess embeddings are replicated across all GPUs, and the accelerator memory stores only the \emph{indices} of the \highaccess embeddings. In a multi-node setup, the learning phase occurs on a single node's \hotline accelerator, after which the indices of \highaccess embeddings are copied to the \hotline accelerators of all nodes.

\niparagraph{2. The Acceleration Phase:} The \hotline system's \emph{acceleration} phase commences once the \highaccess embeddings are replicated on each GPU. During this phase, the \hotline accelerator actively classifies a mini-batch into two $\mu-batches$ based on input popularity. The system employs GPUs to accelerate both $\mu-batches$ through pipelined execution, as depicted in Figure~\ref{fig:hotlinepipe}. Notably, \hotline operates on a finer scale, updating non-\highaccess embeddings on the CPU and \highaccess ones on the GPU\footnote{Prior work, such as FAE~\cite{fae}, have coherence overheads. These overheads stem from the requirement of synchronizing embeddings between CPUs and GPUs, as illustrated in Figure~\ref{fig:latencybreakdown}. These synchronization processes happen at each transition between popular and non-popular mini-batches.}. As a result, \hotline updates embeddings at distinct locations on CPUs or GPUs, thereby avoiding any coherence requests.

The \highaccess embeddings of the popular $\mu$-batch are synchronized across all GPUs with dense parameters via an {\fontfamily{qcr}\selectfont all-reduce} collective. On the other hand, for the non-popular $\mu$-batch, \highaccess embeddings are updated on GPUs, while the remainder is updated on the CPU's main memory using DMA.

\niparagraph{Sources of benefits:} The benefits of \hotline arise from (1) overlapping embedding lookup and communication required for non-popular $\mu-batches$ with GPU-based execution of popular $\mu-batches$, (2) executing all operations, including embedding lookup and update, on GPU HBM using a data and model-aware pipeline scheduler, and (3) using a novel \hotline-accelerator to pipeline segregation and parameter gathering for a single mini-batch without stalling the GPU devices. Roofline analysis showed a theoretical 3$\times$ gain from GPU HBM for embedding lookups over Intel's Optimized Embedding Bag operator~\cite{dlrm-intel} for DDR4 memory. In practice, \hotline achieves nearly a 2.2$\times$ improvement over Intel Optimized DLRM.

\subsection{Model Updates with Hotline}
\label{subsec:hotlineexample}

\begin{figure*}[h]
	\centering
	\includegraphics[width=0.9\textwidth]{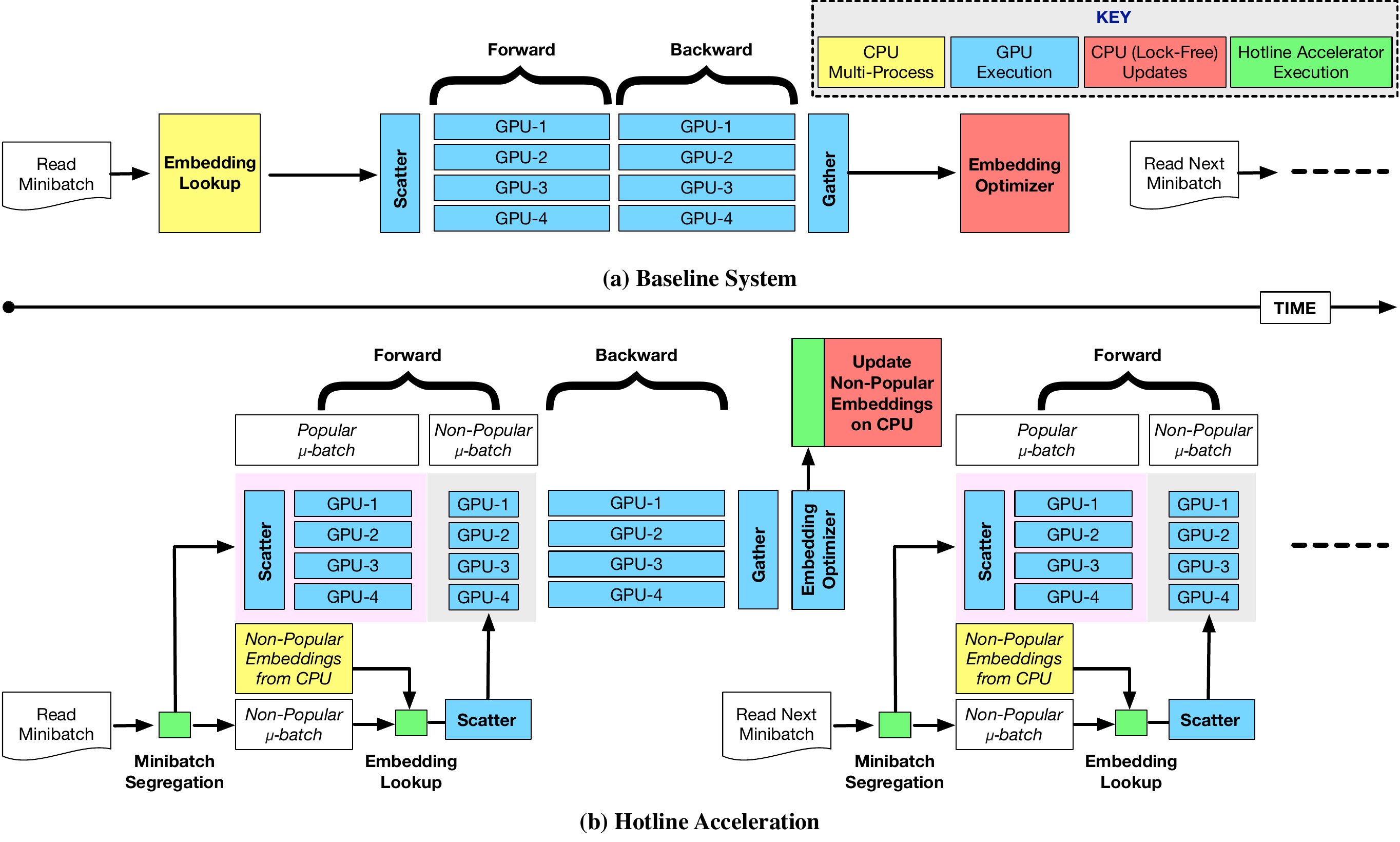}
	\caption{The execution pipeline of \hotline involves the accelerator actively classifying a mini-batch into popular and non-popular $\mu$-batches, then scheduling the popular $\mu$-batch onto the GPU(s). Simultaneously, the accelerator gathers the working parameters for the non-popular $\mu$-batch to schedule onto the GPU(s).}
	\label{fig:hotlinepipe}
\end{figure*}

Click-through rate (CTR) is modeled as a binary classification problem with binary cross-entropy (BCE) loss. A mini-batch ($M$) with $n$ inputs is $ = \{m_1, m_2, \dots, m_n\}$, where $m_i$ is a single input. Across mini-batch $M$, the BCE loss ($L$) for each input $f(m_i)$ is represented as Equation~\ref{eqn:1}, where $y_i$ is the target and $p_i$ is the predicted probability for the $i^{th}$ input.
\setlength{\belowdisplayskip}{3pt} \setlength{\belowdisplayshortskip}{3pt}
\setlength{\abovedisplayskip}{3pt} \setlength{\abovedisplayshortskip}{3pt}
\begin{equation}\label{eqn:1}
    f(m_i) = y_i\log(p_i) + (1 - y_i)\log(1 - p_i)
\end{equation}
\noindent Thus the BCE loss ($L$) for $M$ is represented as Equation~\ref{eqn:2}.
\begin{equation}\label{eqn:2}
\begin{aligned}
    L_{baseline} = \sum_{i=1}^{n} f(m_i) = \sum_{i=1}^{n} y_i\log(p_i) + (1 - y_i)\log(1 - p_i)
\end{aligned}
\end{equation}
\noindent As shown in Figure~\ref{fig:hotlinepipe}, \hotline splits mini-batch $M$ into two $\mu$-batches: popular and non-popular. The popular $\mu$-batch is represented as $\mathcal{O} = \{o_1, o_2, \dots, o_l\}$. Similarly, the non-popular $\mu$-batch is represented as $\mathcal{X} = \{x_1, x_2, \dots, x_k\}$. 
Often $l > k$ because the popular inputs are a larger portion of the dataset.
The two $\mu$-batches are mutually exclusive, i.e., without overlapping inputs. We express this using Equation~\ref{eqn:3}.
\begin{equation}\label{eqn:3}  
\begin{aligned}
    \mathcal{O} \cup \mathcal{X} = M \hspace{0.3in} 
    \mathcal{O} \cap \mathcal{X} = \emptyset 
\end{aligned}
\end{equation}
\noindent The BCE loss for $\mathcal{O}$ and $\mathcal{X}$ is denoted by Equation~\ref{eqn:4}:
\begin{equation}\label{eqn:4}
\begin{aligned}
    L_{popular} &= \sum_{i=1}^{l} f(o_i) \\
    L_{non-popular} &= \sum_{i=1}^{k} f(x_i)
\end{aligned}
\end{equation}
The BCE loss of \hotline is denoted as $L_{hotline} = L_{popular} + L_{non-popular}$. Now, using Equation~\ref{eqn:3}, and Equation~\ref{eqn:4}, we can rewrite Equation~\ref{eqn:2} as Equation~\ref{eqn:5}.
\begin{equation}\label{eqn:5}
\begin{aligned}
    L_{baseline} &= \sum_{i=1}^{l} f(o_i) + \sum_{i=1}^{k} f(x_i) = \sum_{i=1}^{n} f(m_i)\\
                   &= L_{popular} + L_{non-popular} \\
                   &= L_{hotline}
\end{aligned}
\end{equation}
Therefore, the BCE loss calculated for baseline and \hotline is the same. Consequently, their gradients during back-propagation are also identical. Thus, compared to baseline, \hotline depicts \emph{no loss} in training or testing accuracy.

\section{The Hotline Accelerator}
\label{sec:architecture}
\begin{figure}[h!]
	\centering
	\includegraphics[width=0.85\columnwidth]{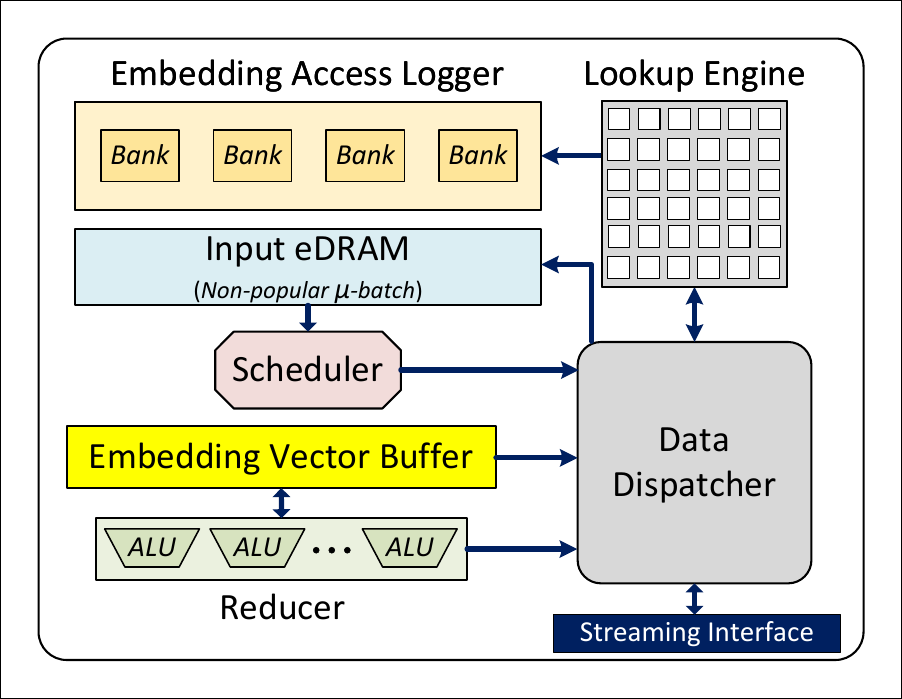}
	\caption{The components within the \hotline hardware accelerator block. The \hotline accelerator is connected to the low-profile PCIe slot via a Streaming Interface.}
	\label{fig:acceleratorflow}
\end{figure}

Figure~\ref{fig:acceleratorflow} shows the block diagram of the \hotline accelerator. We will now describe the micro-architectural details of each component within the \hotline hardware accelerator. 

\subsection{Data Dispatcher}
\begin{figure}[h!]
	\centering
	\includegraphics[width=1\columnwidth]{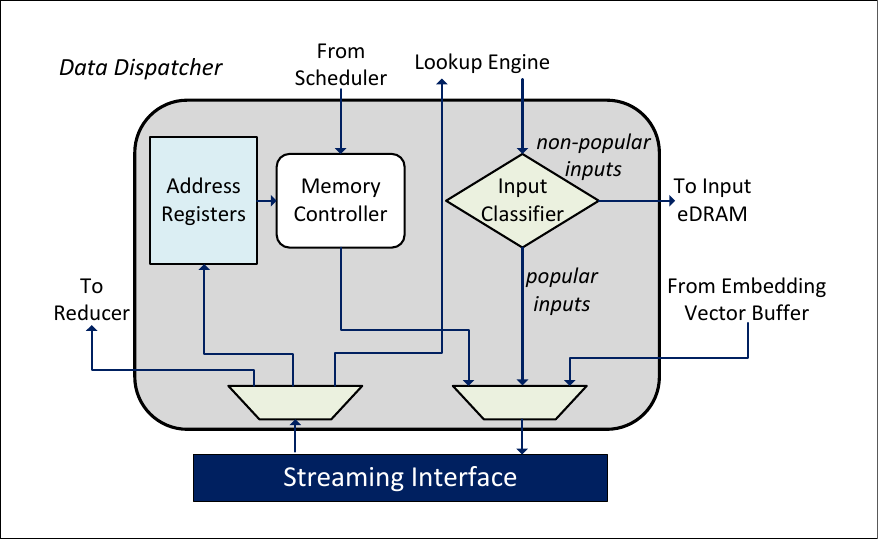}
	\caption{The Address Registers, Memory Controller, and Input Classifier constitute the Data Dispatcher block.}
	\label{fig:datadispatcher}
\end{figure}

Figure~\ref{fig:datadispatcher} shows the Data Dispatcher block, which includes the \emph{Address Registers} containing the base address of each embedding table in CPU and GPU memory. The \emph{Memory Controller} uses these Address Registers to generate embedding addresses. The \emph{Input Classifier} segregates incoming inputs based on the Embedding Access Logger (EAL) information, distinguishing them as popular or non-popular. While the popular $\mu$-batch executes, the dispatcher sends the non-popular $\mu$-batch from the input eDRAM to the Lookup Engine. Our design shows that a small 2.5~MB of eDRAM can store mini-batches with up to 16K inputs.

The non-popular $\mu$-batch accesses arbitrary embeddings. The memory controller sends a direct memory access (DMA) request to the DMA engine for not-\highaccess embeddings and initiates a GPU read memory request for \highaccess embeddings. The Reducer block processes working parameters from CPU and GPU memory, reducing multiple embedding rows into a single embedding vector, and then stores it in the \emph{Embedding Vector Buffer}.

\subsection{Embedding Access Logger (EAL)}
\label{subsec:EAL}

\begin{figure}[h!]
	\centering
	\includegraphics[width=1\columnwidth]{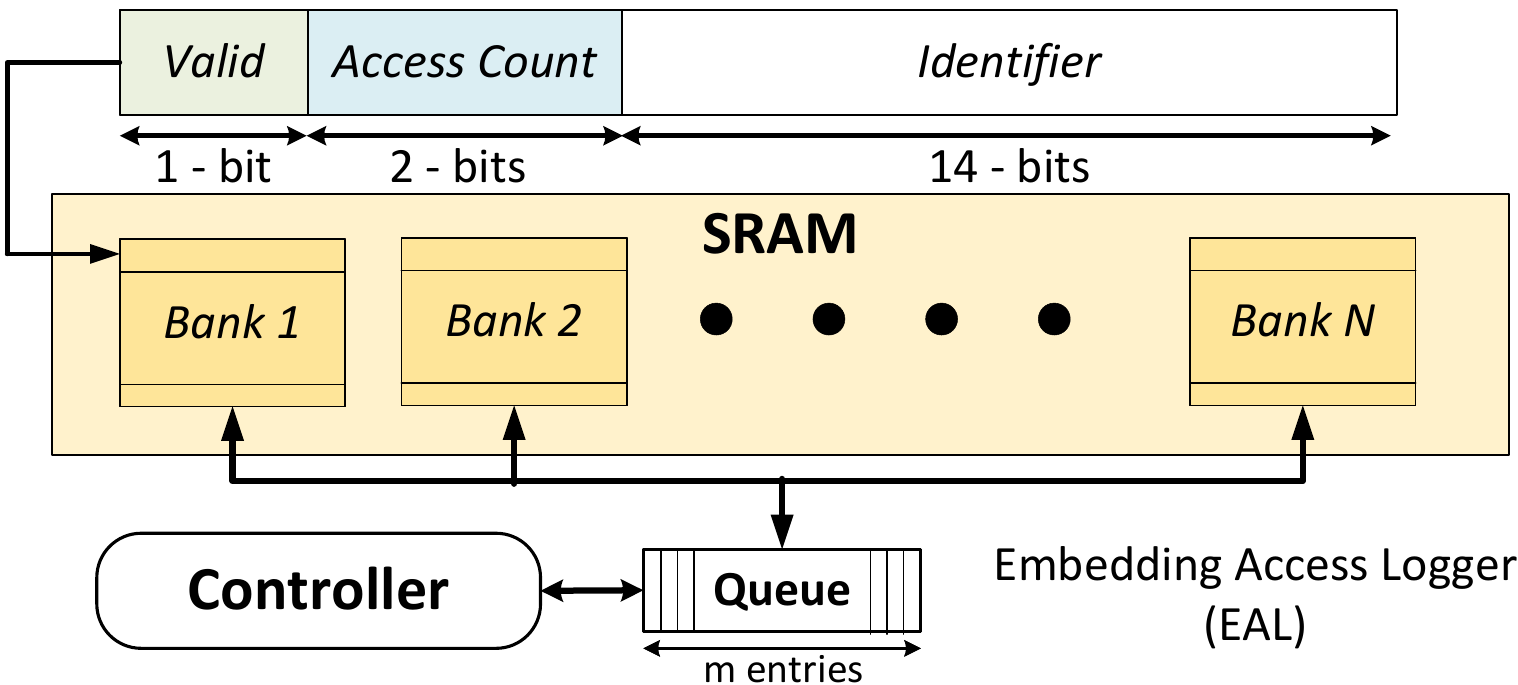}
	\caption{The Embedding Access Logger (EAL) block consists of Multi-Banked SRAM, the Controller, and the Queue sub-blocks. It tracks frequently accessed embeddings.}
	\label{fig:EALentry}
\end{figure}

The Embedding Access Logger (EAL) actively utilizes a counter to track the frequency of access to embedding entries. EAL stores only the indices of embedding entries with valid bits and access counts. Figure~\ref{fig:EALentry} depicts the components of EAL, which include a multi-banked Static Random Access Memory (SRAM), a controller, and a queue.

\niparagraph{A. Naive Embedding Tracking:} Due to a large number of sparse parameters in recommender models, per-entry frequency counters would require gigabytes of on-chip storage. Alternatively, storing the frequencies in CPU/GPU memory would require three accesses: one to obtain the embeddings, one to read the frequency, and another to update it.

\niparagraph{B. Efficient Embedding Tracking:} The EAL design is motivated by two observations: the size of \highaccess embeddings is $\leq$512 MB and their access skews are extremely high. Thus, EAL is designed as a cache-like structure that tracks \highaccess embedding indices using a 4~MB SRAM cache with 2 million blocks. EAL uses the Static Re-reference Interval Predictor (SRRIP) replacement policy with a 2-bit Re-reference Predictor Value (RRPV) counter to reduce area overheads. As \highaccess embeddings have $>$100$\times$ more frequent accesses, a 2-bit RRPV counter (access counter or AC) with insertions at RRPV-1 value captures $>$99\% of the \highaccess embeddings with 70\% tracking capability. Even if a non-\highaccess embedding is misclassified as \highaccess or vice versa, it has no impact on model fidelity.

\begin{figure}[t!]
	\centering
	\includegraphics[width=0.9\columnwidth]{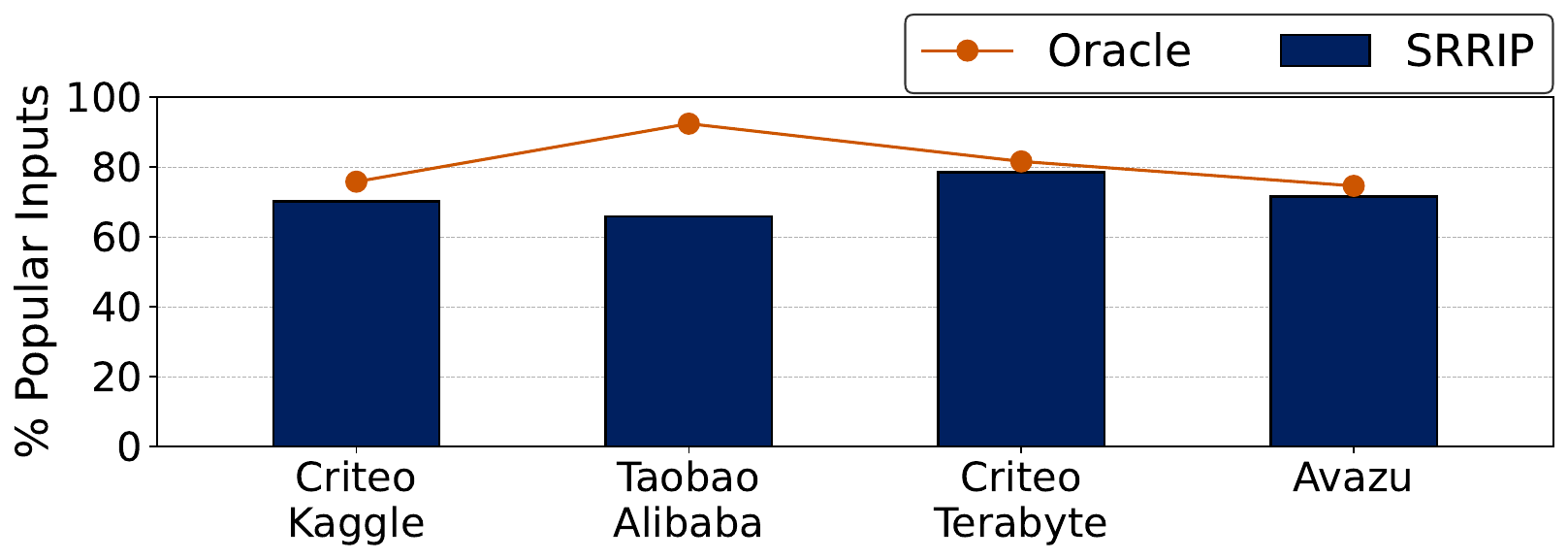}
	\caption{SRRIP-based tracker as compared to the Oracle LFU scheme. On average, the SRRIP-based tracker can track 90\% of the \highaccess embeddings.}
	\label{fig:ip_hotness_roofline}
	\vspace{-0.1in}
\end{figure}

To evaluate EAL with one-hot encoded inputs versus multi-hot encoded inputs, we compared the hit rate of one-hot encoded real-world datasets with multi-hot encoded synthetic datasets~(Section~\ref{subsubsec:syn_models}). The hit rate of EAL for the multi-hot encoded datasets decreases by only a maximum of 5\%. Figure~\ref{fig:ip_hotness_roofline} compares the SRRIP logger to an Oracle logger~\footnote{We could use the Least Frequently Used (LFU) replacement policy to understand the access frequency. However, this incurs significant area overheads, as each cache block would require a 24-bit counter (Figure~\ref{fig:ip_emb_acc} shows embeddings can have up to 10 million accesses).}.

\begin{figure}[h!]
	\centering
	\includegraphics[width=0.9\columnwidth]{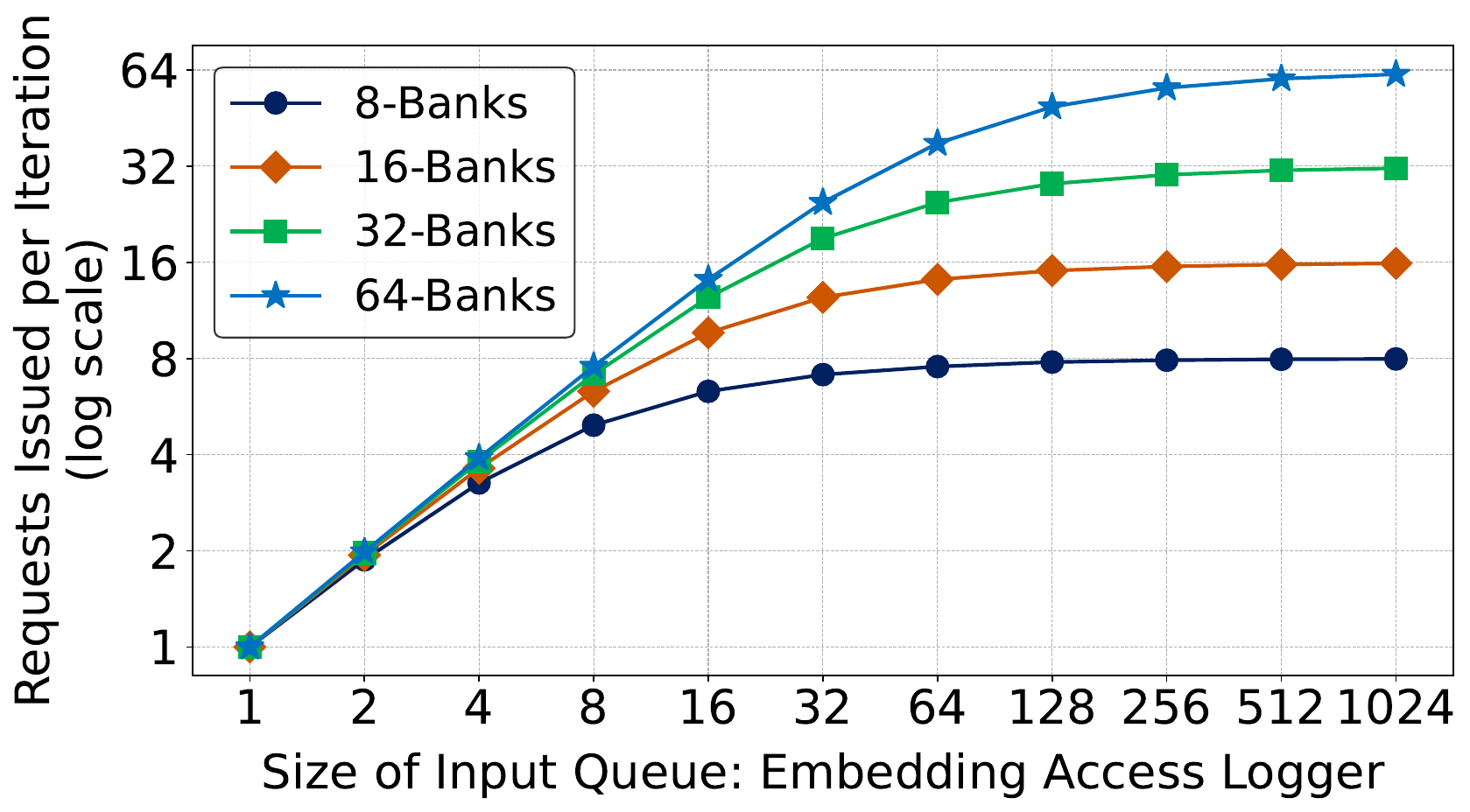}
	\caption{Impact of Queue Size and Banks on the number of parallel requests per iteration. Embedding Access Logger uses a 512 queue with 64 banks to enable 60 parallel requests.}
	\label{fig:queuesize}
\end{figure}

\niparagraph{C. Multi-Banked SRAM for Parallel Lookup:} \hotline enables parallel lookups by dividing the EAL into multiple banks. Figure~\ref{fig:queuesize} shows our empirical design space exploration, which reveals the average number of requests issued as the number of banks ($n$) and input queue size ($m$) vary. On average, a 512-sized queue with 64 banks allows for 60 parallel requests per iteration without collisions. A controller schedules requests from the lookup engine block to the 512-entry queue. Periodically, the EAL switches to the `learning' phase to capture changes in popular embeddings.

\subsection{The Lookup Engine}
The Lookup Engine is a parallel 2D lookup network that extracts embedding entries from every training input. It can achieve 26$\times$ throughput per input if it requires 26 distinct embedding tables. Additionally, the 2D lookup network allows for exploiting parallelism within the mini-batch. During the learning phase, the Lookup engine provides EAL with the indices accessed by each input. The Lookup Engine classifies inputs as popular during the acceleration phase if all embedding indices are within EAL.

\begin{figure}[t!]
	\centering
	\includegraphics[width=\columnwidth]{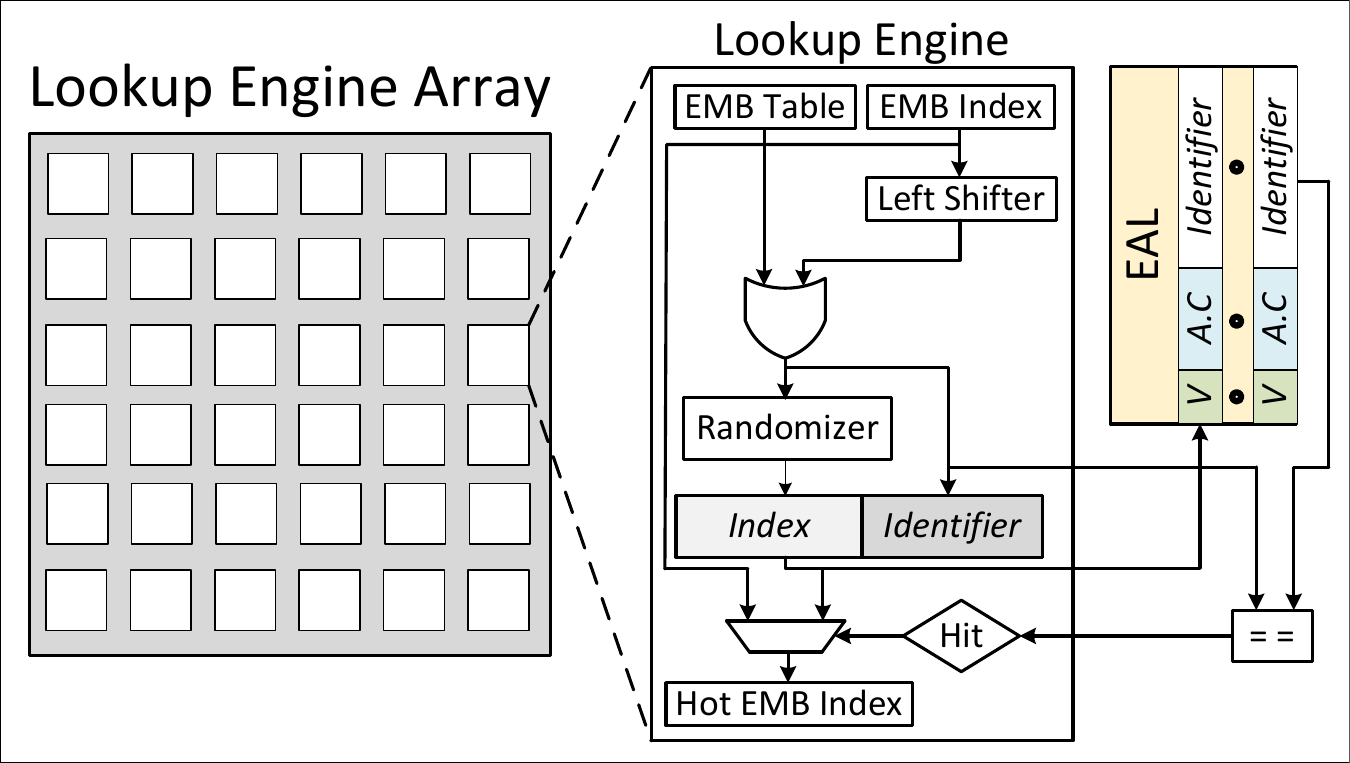}
	\caption{The Lookup Engine. The lookup engine determines the embedding entries required by an input.}
	\label{fig:pe}
\end{figure}

A single lookup engine, as shown in Figure~\ref{fig:pe}, contains registers for embedding table numbers, hot embedding index, and a randomizer. The randomizer hashes the (Embedding Index, Embedding Table) tuple to scatter embedding index values across EAL and prevent trashing. A low-latency Fiestal Network implements the randomizer~\cite{fiestal}.

\subsection{The Reducer}
The Reducer performs a sparse-length element-wise sum operation using a simple arithmetic unit array. It reduces multiple embedding rows into a single vector through a pooling operation and saves the result in the Embedding Vector Buffer.

\subsection{Instruction Set Architecture}
\label{subsec:hotline_driver}
The \hotline accelerator relies on a driver to communicate with the CPU's main memory and GPU devices. The driver interacts with the DMA engine to access not-\highaccess embeddings on CPU main memory and GPU devices via a PCIe link. It uses instructions, as listed in Table~\ref{table:isa}, to read/write the necessary data into these devices.

\begin{table}[h!]
\centering
\caption{\hotline's Instruction Set}
\resizebox{1\columnwidth}{!}{
\begin{tabular}{ l c c l}
 \hline
 \textbf{Instruction} & \textbf{Operand 1} & \textbf{Operand 2} & \textbf{Description}  \\
 \hline
 dma\_rd(op1, op2) & mem start idx & \# bytes & DMA read request  \\
 dma\_wr(op1, op2) & mem start idx & \# bytes & DMA write request  \\
 v\_add(op1, op2) & input vector & emb vec buff & element wise addition  \\
 v\_mul(op1, op2) & input vector & emb vec buff & element wise dot product  \\
 s\_wr(op1, op2) & reg idx & base addr & write emb base addr  \\
 gpu\_rd(op1, op2) & gpu device id & sparse idx & read emb idx from GPU device \\
 \hline

\end{tabular}}
\label{table:isa}
\end{table}

\begin{table*}
\centering
\caption{Recommender Model Architecture and Parameters}
\newcolumntype{?}{!{\vrule width 2pt}}
\setlength\extrarowheight{4pt}
\resizebox{0.9\textwidth}{!}{
\begin{tabular}{c | l | c | c c | c c c | c c c | c }
\hline
\multirow{2}{*}{\textbf{Model}} & \multirow{2}{*}{\textbf{Dataset}} &
\textbf{Time} &
\multicolumn{2}{c|}{\textbf{Features}} & \multicolumn{3}{c|}{\textbf{Parameters}} &  \multicolumn{3}{c|}{\textbf{Neural Network Configuration}} &
\multirow{2}{*}{\textbf{Size}}
\\
\cline{4-11} &  & \textbf{Series} & \textbf{Dense} & \textbf{Sparse} & \textbf{Dense} & \textbf{Sparse} & \textbf{Sparse Dim} & \textbf{Bottom MLP} & \textbf{Top MLP} & \textbf{DNN} & \textbf{(GB)} \\
\hline
\bench{RM1} & Taobao Alibaba~\cite{alibaba} & 21 & 1 & 3 & 7.3~k & 5.1~M & 16 & 1-16 & 30-60-1 & Attn. Layer & 0.3 \\
\bench{RM2} & Criteo Kaggle~\cite{criteokaggle} & 1 & 13 & 26 & 287.5~k & 33.8~M & 16 & 13-512-256-64-16 & 512-256-1 & - & 2 \\
\bench{RM3} & Criteo Terabyte~\cite{criteoterabyte} & 1 & 13 & 26 & 549.1~k & 266~M & 64 & 13-512-256-64 & 512-512-256-1 & - & 63 \\
\bench{RM4} & Avazu~\cite{avazu} & 1 & 1 & 21 & 281.4~k & 9.3~M & 16 & 1-512-256-64-16 & 512-256-1 & - & 0.55 \\
 \hline
\end{tabular}}
\label{table:datasets}
\end{table*}

\section{Evaluation Methodology}
\label{sec:eval}

\begin{figure*}[t]
  \centering
  \subfloat[Criteo Kaggle]{
	\begin{minipage}[t]{0.23\textwidth}
	   \centering
	   \includegraphics[width=\textwidth]{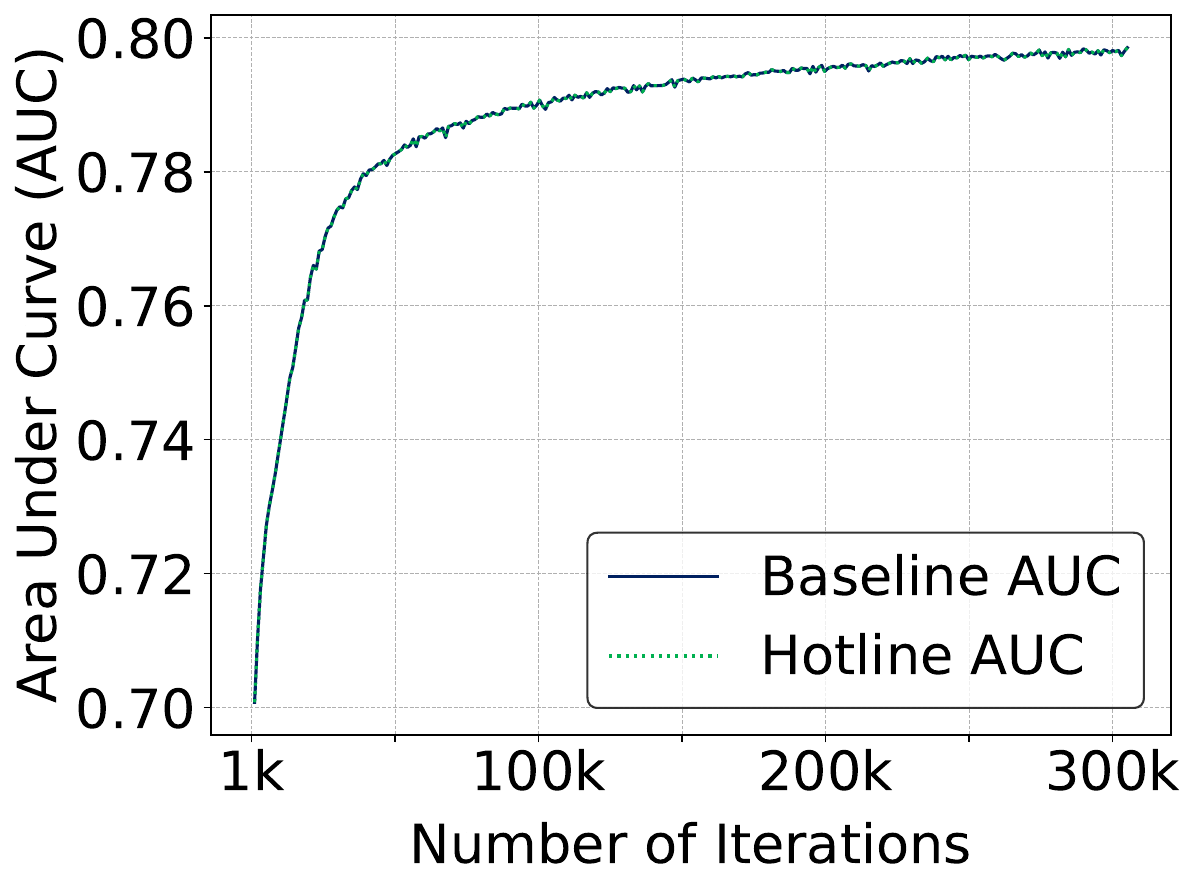}
	\end{minipage}}
  \subfloat[Taobao Alibaba]{
	\begin{minipage}[t]{0.23\textwidth}
	   \centering
	   \includegraphics[width=\textwidth]{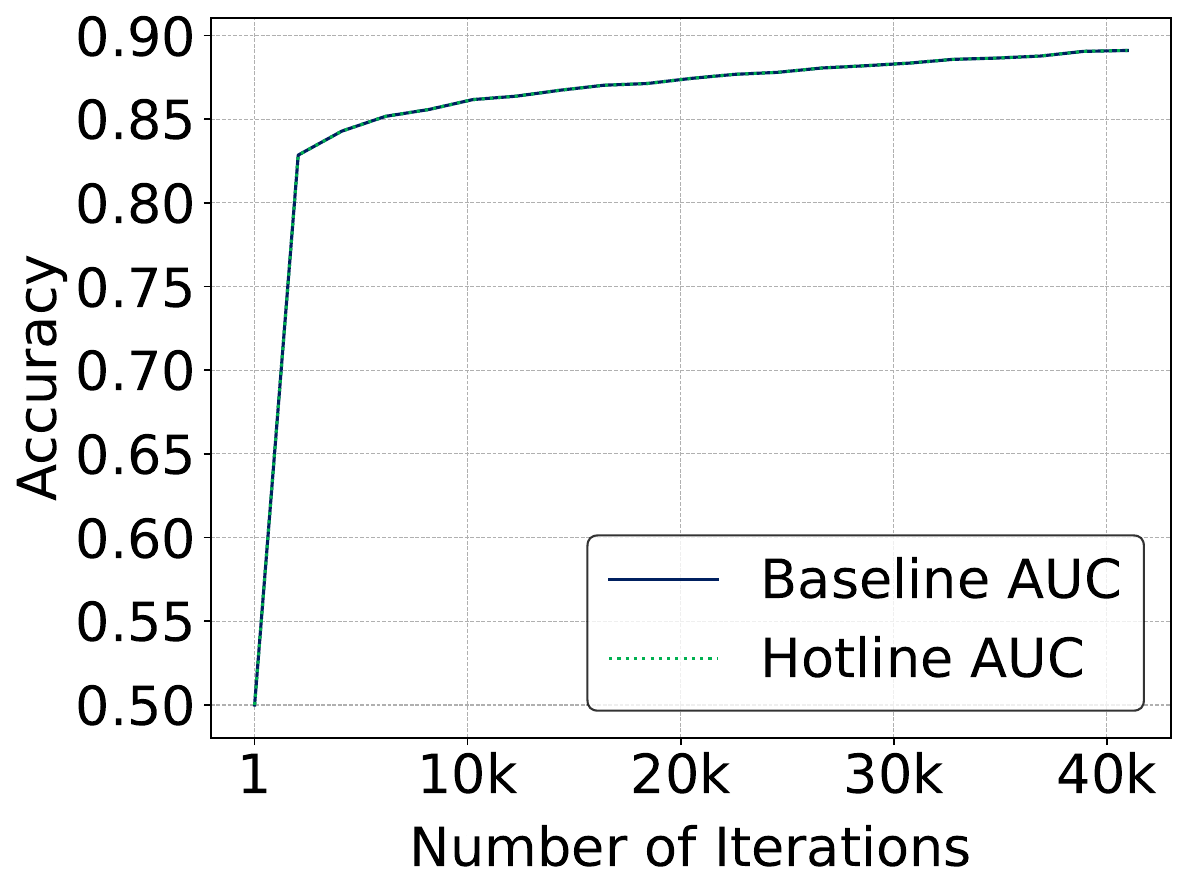}
	\end{minipage}}
  \subfloat[Criteo Terabyte]{
	\begin{minipage}[t]{0.23\textwidth}
	   \centering
	   \includegraphics[width=\textwidth]{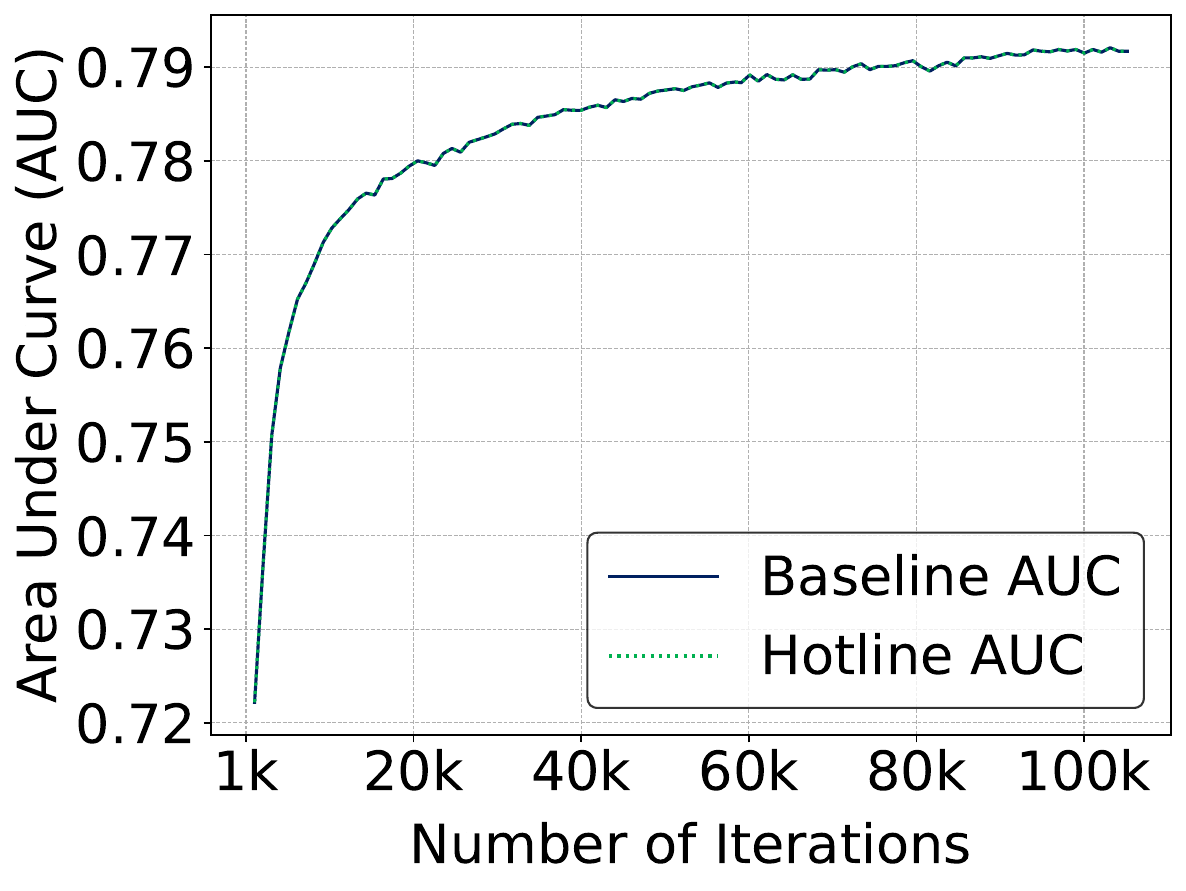}
	\end{minipage}}
  \subfloat[Avazu]{
	\begin{minipage}[t]{0.23\textwidth}
	   \centering
	   \includegraphics[width=1\textwidth]{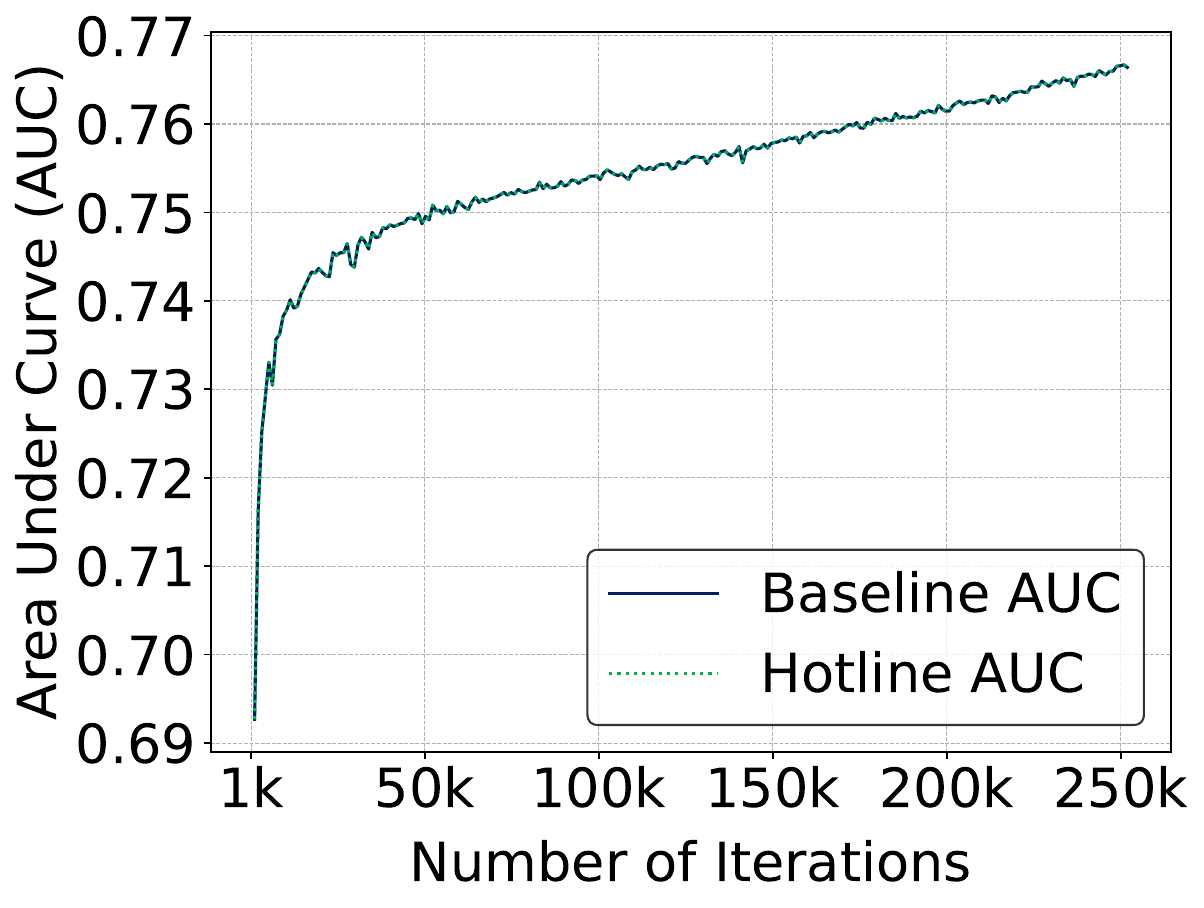}
	\end{minipage}}
\caption{The accuracy of \hotline with full-precision training. \hotline maintains exactly identical training fidelity as the baseline.}
\label{fig:accuracy}
\end{figure*}

\subsection{Models}
\label{subsec:models}
Table~\ref{table:datasets} presents the specifications of four open-sourced recommender models that were evaluated using \hotline. The models have varying numbers of sparse parameters, ranging from 5.1M for \bench{RM1} to 266M for \bench{RM3}. These models consist of a top and bottom multi-layer perceptron (MLP) with a deep learning attention layer for \bench{RM1}. On the other hand, \bench{RM2} and \bench{RM3} have more sparse features and larger embedding tables, making them embedding-dominated models. \bench{RM4} has an average-sized dense neural network and sparse embedding tables. Benchmarks such as Deep Learning (DLRM)~\cite{dlrm} and Time-based Sequence Models (TBSM)~\cite{tbsm} were used to train these models, with TBSM training the \bench{RM1} model and DLRM training the \bench{RM2}, \bench{RM3}, and \bench{RM4} models.

\subsection{Datasets}
We train on four real-world datasets, listed in Table~\ref{table:datasets}. Taobao Alibaba~\cite{alibaba} is a user behavior dataset for recommendation problems with implicit feedback. Criteo Kaggle~\cite{criteokaggle} dataset contains advertising data and is obtained from the Display Advertising Challenge to capture user preferences by predicting CTR. Criteo Terabyte~\cite{criteoterabyte} is the largest publicly available dataset for user click logs. The Avazu~\cite{avazu} dataset is taken from a CTR prediction competition by Kaggle.

\subsection{Software libraries and setup}
\label{subsec:sfotware_setup} 
We configured DLRM and TBSM using Pytorch-1.9 with the \emph{torch.distributed} backend to support scalable distributed training and performance optimizations~\cite{pytorch}. To achieve GPU-to-GPU communication for collective operations like gather, scatter, and all-reduce, we used NVIDIA Collective Communication Library (NCCL)~\cite{nccl} on NVLink-2.0. We compared our results with other implementations such as XDL~\cite{xdl}, Intel-optimized DLRM~\cite{dlrm-intel}, and FAE~\cite{fae}. The XDL-based implementation uses Tensorflow-1.2~\cite{tensorflow}.

\begin{table}[!h]
\centering
\caption{System Specifications}
\vspace{-0.05in}
\resizebox{1\columnwidth}{!}{
\begin{tabular}{c c c c}
 \hline
 \textbf{Device} & \textbf{Architecture} & \textbf{Memory} & \textbf{Storage}  \\
 \hline
 CPU & Intel Xeon & 192~GB  & 1.9~TB \\
  & Silver 4116 (2.1~GHz) & DDR4 (76.8~GB/s) & NVMe SSD \\
  \hline
 GPU & Nvidia Tesla & 16~GB  &  - \\
     & V100 (1.2~GHz)& HBM-2.0 (900~GB/s) \\
 \hline
\end{tabular}}
\label{tab:systemspecs}
\end{table}

\subsection{Server Specifications}

Table~\ref{tab:systemspecs} provides information about the server used for the experiments. The server employs a 24-core Intel Xeon Silver 4116 (2.1 GHz) processor based on Skylake architecture and is equipped with 4 NVIDIA Tesla-V100 GPUs. The communication between the GPUs, \hotline accelerator, and the rest of the system is facilitated via a 16x PCIe Gen3 bus. All experiments are conducted on a single server.

\subsection{Measurements} 
\label{subsec:measurements}

We measure the model's convergence time using wall clock time. The Verilog RTL architecture of the \hotline accelerator is validated using Synopsys DC at 350 MHz with 45$nm$ technology. Cacti is used to estimate the area/energy of memory components and their access time. The accelerator details can be found in Table~\ref{tab:accelspecs}. The runtime of the accelerator is calculated using the compute and access cycles obtained from Synopsys DC and Cacti through RTL simulation. Additionally, we estimate the time it takes to gather the working parameters using real-system DMA and HBM latencies and incorporate this latency in the pipeline. The end-to-end training time includes the latency for executing the non-popular $\mu$-batch with parameters already available on the GPU. To mitigate HBM contention, we fetch \highaccess embeddings from different GPUs in a round-robin fashion, ensuring a balanced memory load on each device.

\begin{table}[!h]
\vspace{-0.1in}
\centering
\caption{Accelerator Specifications}
\vspace{-0.05in}
\resizebox{\columnwidth}{!}{
\begin{tabular}{ l c|l c}
 \hline
 \textbf{Parameters} & \textbf{Settings} & \textbf{Parameters} & \textbf{Settings}  \\
 \hline
 Frequency & 350~MHz &  EAL size & 4~MB \\
 No of Reducer ALU Units &  16 & No of Lookup Engines &  64 \\
 Input eDRAM size & 2.5~MB & Embedding Vector Buffer & 0.5~kB \\
 Total Area & 7.01~$mm^2$ &  Average Energy & 132 mJ \\
 \hline
\end{tabular}}
\label{tab:accelspecs}
\end{table}

\begin{figure*}
	\centering
	\includegraphics[width=0.95\textwidth]{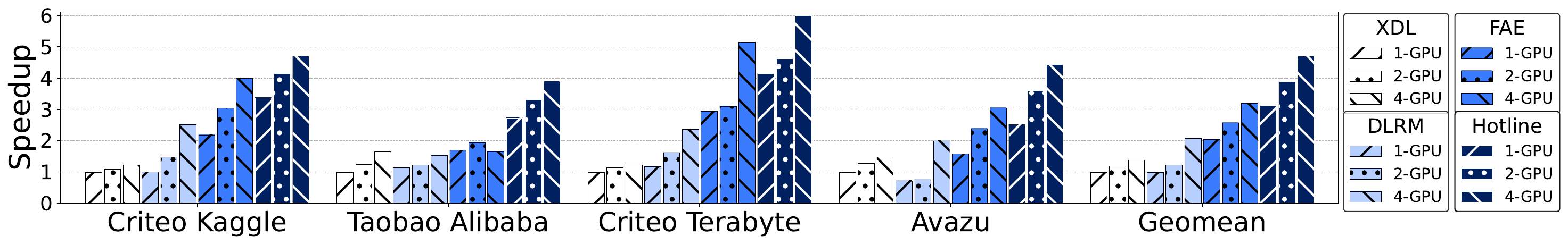}
	\caption{The performance comparison of \hotline with XDL, Intel optimized DLRM, and FAE implementations (normalized to a 1-GPU XDL). On average, even a 1-GPU \hotline provides \avgperfimpronegpudlrm higher performance than the baseline.}
	\label{fig:performance}
\end{figure*}

\begin{figure*}
	\centering
 \includegraphics[width=0.98\textwidth]{./body/Graphs/perf_breakdown/perf_breakdown.png}
	\caption{Latency breakdown of 1, 2, and 4 GPU implementations of software frameworks and \hotline. The overhead of CPU-GPU communication time increases as the number of GPUs scale because of inter-GPU communication. }
	\label{fig:latencybreakdown}
\end{figure*}

\section{Results and Analysis}
\subsection{Training Accuracy}
\label{sec:accuracy}

We evaluated the accuracy of \hotline using full-precision DLRM and TBSM model implementations. Figure~\ref{fig:accuracy} illustrates the Area Under Curve (AUC) accuracy metric for Kaggle, Terabyte, and Avazu established by MLPerf~\cite{mlperf, mlperf_rec}. We observed that \hotline followed the baseline test and train accuracy and had \emph{no accuracy implications}. This is because \hotline fragments a mini-batch into two $\mu$-batches that continue to update the same embeddings. The baseline implementation and \hotline update these embeddings with identical gradients in each mini-batch. Table~\ref{table:accuracy} also compares the testing accuracy, AUC, and cross-entropy loss across datasets for DLRM (baseline) and \hotline.

\begin{table}[h!]
\centering
\caption{Comparison of Accuracy Metrics}
\scriptsize
\newcolumntype{?}{!{\vrule width 2pt}}
\newcolumntype{P}[1]{>{\centering\arraybackslash}p{#1}}
\setlength\extrarowheight{4pt}
\resizebox{\columnwidth}{!}{
\begin{tabular}{ l | c c c | c c c }
\hline
\multirow{2}{*}{\textbf{Dataset}} &  \multicolumn{3}{c}{\textbf{DLRM}} & \multicolumn{3}{|c}{\textbf{\hotline}}
\\
\cline{2-7} & \textbf{Accuracy (\%)} & \textbf{AUC} & \textbf{Logloss} & \textbf{Accuracy (\%)} & \textbf{AUC} & \textbf{Logloss}
\\
\hline
Criteo Kaggle & 78.64 & 0.798 & 0.456 & 78.64 & 0.798 & 0.456\\
 Taobao Alibaba & 89.11 & 92.61 & 0.270 & 89.11 & 92.61 & 0.270 \\
 Criteo Terabyte & 81.20 & 0.792 & 0.421 & 81.20 & 0.792 & 0.421 \\
 Avazu & 83.61 & 0.766 & 0.387 & 83.61 & 0.766 & 0.387\\
 \hline

\end{tabular}}
\label{table:accuracy}
\end{table}

\subsection{Comparison with Hybrid Baseline}
\label{subsec:hybrid_baseline_comparison}

\subsubsection{Performance comparisons}
Figure~\ref{fig:performance} compares \hotline with three state-of-the-art software implementations while varying the number of GPUs. XDL~\cite{xdl} uses a parameter server approach, Intel-Optimized DLRM~\cite{dlrm-intel} executes embeddings on the CPU with lock-free updates, and FAE~\cite{fae} utilizes input popularity with \textit{offline pre-processing} and \textit{CPU-based scheduling without pipelining}.
While \hotline is designed to adapt to changing trends in user behaviour (Section~\ref{subsec:challenge3} (Challenge 3) and Figure~\ref{fig:evolving_skew}), FAE cannot capture such changing trends as it employs a static offline profiler while also incurring a 15\% overhead. In contrast, \hotline periodically updates the frequently accessed embeddings with minimal overhead.
We use weak scaling to scale mini-batch size with GPUs, and all numbers are normalized to XDL's 1-GPU setup.

Figure~\ref{fig:performance} shows that \hotline reduces training time as the recommender model is executed on GPU(s). \hotline has a speedup of \avgperfimpronegpuxdl and \avgperfimprtwogpuxdl for 1-GPU and 2-GPU setups, and \avgperfimprfourgpuxdl for 4-GPU setups, on average across all models and datasets, over XDL. Compared to optimized DLRM, \hotline has a speedup of \avgperfimpronegpudlrm, \avgperfimprtwogpudlrm, and \avgperfimprfourgpudlrm for 1-GPU, 2-GPU, and 4-GPU implementations, respectively. \hotline also outperforms FAE with a speedup of \avgperfimpronegpufae, \avgperfimprtwogpufae, and \avgperfimprfourgpufae for the 1-GPU, 2-GPU, and 4-GPU setups, respectively. This is due to efficient runtime scheduling on a massively-parallel \hotline accelerator instead of the CPU.

\subsubsection{Latency breakdown}

Figure~\ref{fig:latencybreakdown} demonstrates the latency breakdown of \hotline and three hybrid baselines. The Criteo Kaggle and Terabyte datasets, which are more embedding and memory intensive, comprise high CPU--GPU communication time. \hotline eliminates the CPU-GPU communication time for popular $\mu$-batch being completely executed on GPU. In contrast, for non-popular $\mu$-batch, it hides the parameter gathering under popular $\mu$-batch execution. In the case of the Taobao dataset, which is dominated by the neural network, deep learning execution surpasses the communication time.
Overall overhead is shown in Figure~\ref{fig:latencybreakdown}. 
This overhead for \hotline includes online profiling and is minimal, primarily because online profiling done at the start of training is not hidden under GPU execution. Still, all subsequent profiling is hidden under GPU execution, significantly reducing overhead. Also, the lookup engine parallelizes the input accesses from EAL for embedding indices. This results in fast online profiling. In our evaluation, we transitioned to the access learning phase twice within a single epoch. Users can specify the frequency at which the learning phase is invoked. 

In contrast, our experiments reveal that offline profilers have a 15\% additional overhead in training time~\cite{fae}. Prior work often overlooks this overhead of the static offline profiler in the overall training time~\cite{fae, recshard}. Additionally, FAE incurs coherence overhead from embedding synchronization when switching between popular and non-popular data.

\subsubsection{Throughput improvements}
\begin{figure}[t!]
  \centering
	\begin{minipage}[t]{0.255\linewidth}
	   \centering
	   \includegraphics[width=\textwidth]{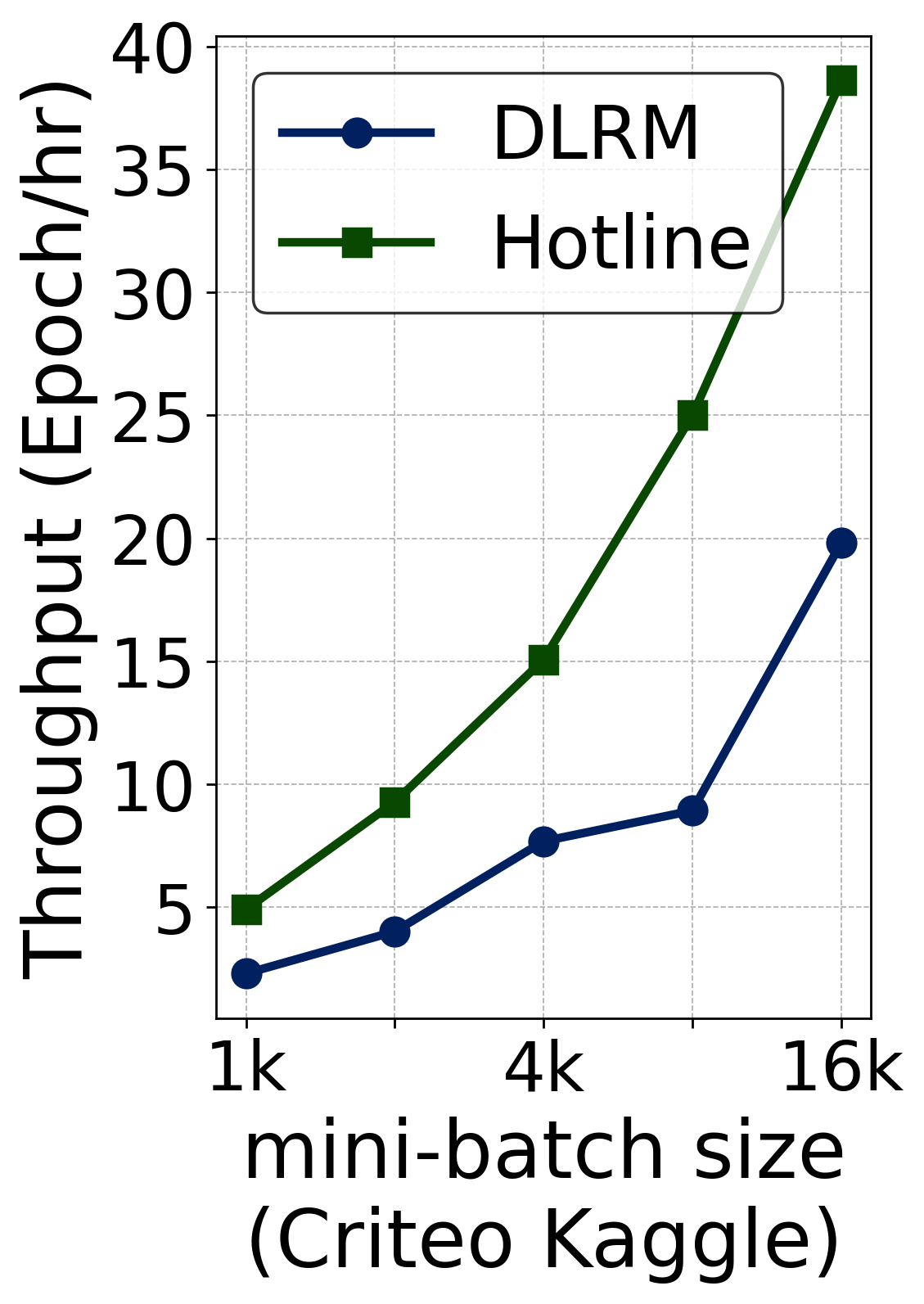}
	\end{minipage}
	\begin{minipage}[t]{0.2325\linewidth}
	   \centering
	   \includegraphics[width=\textwidth]{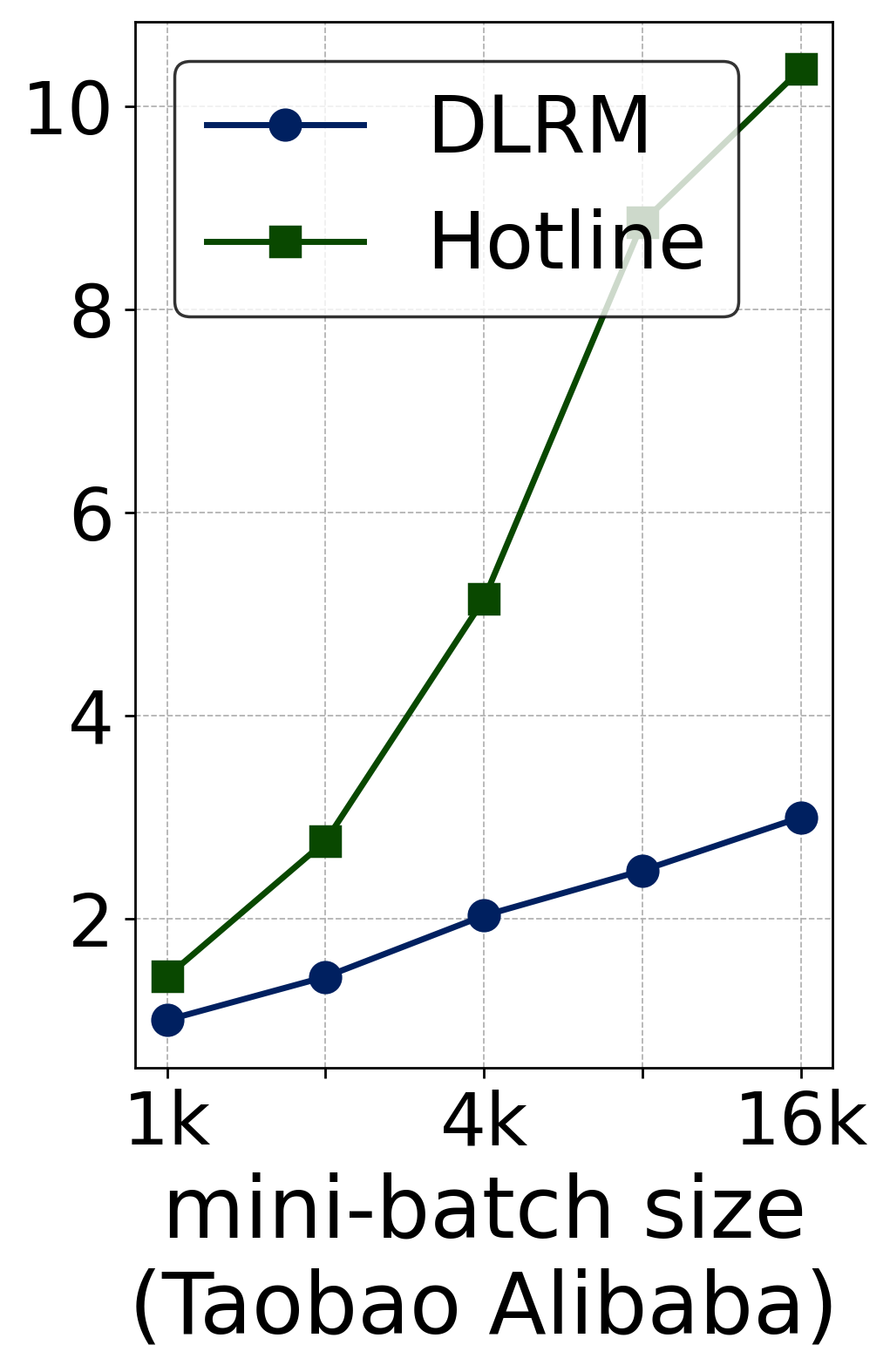}
	\end{minipage}
	\begin{minipage}[t]{0.2325\linewidth}
	   \centering
	   \includegraphics[width=\textwidth]{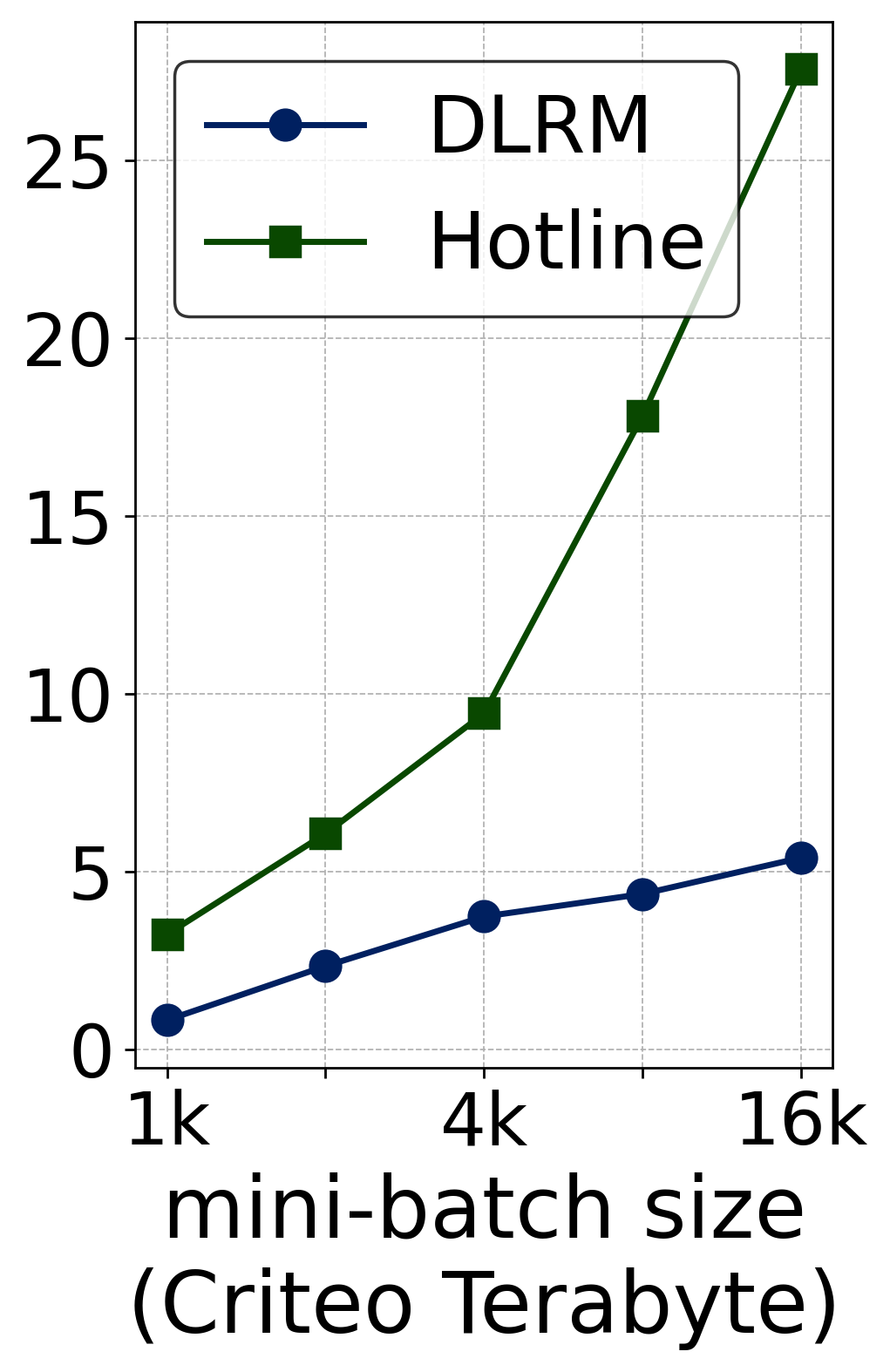}
	\end{minipage}
	\begin{minipage}[t]{0.235\linewidth}
	   \centering
	   \includegraphics[width=\textwidth]{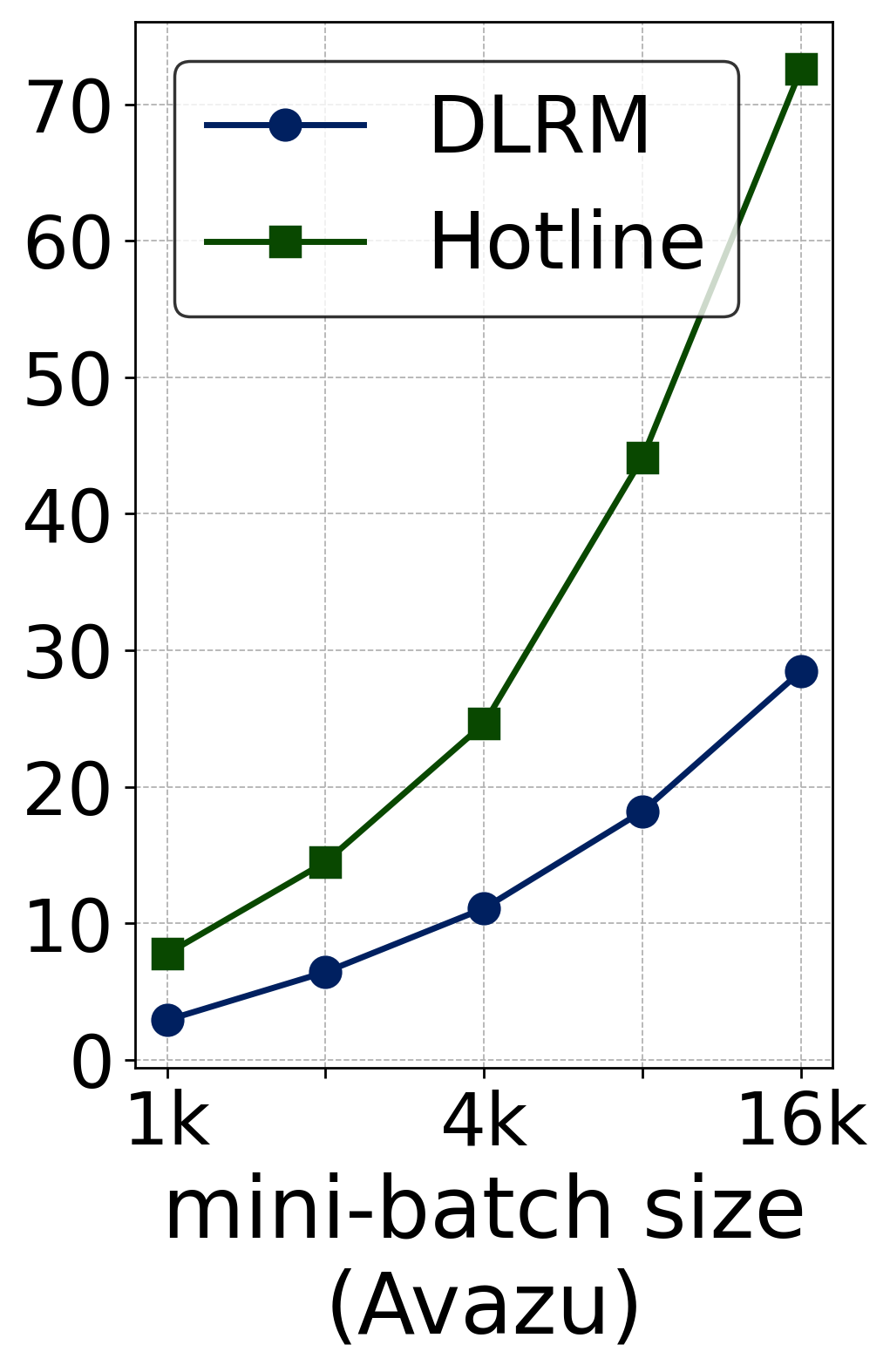}
	\end{minipage}
\caption{Training throughput (Epochs/hour) with 4-GPU executions. Hotline achieves, on average, 2.6$\times$ more throughput than the optimized DLRM baseline.}
\label{fig:throughput}
\end{figure}

Figure~\ref{fig:throughput} shows that for a 4-GPU system, \hotline achieves higher throughput than the optimized DLRM baseline, averaging \avgthroughputimp more epochs/hour. The throughput of \hotline increases rapidly for larger mini-batches due to its ability to utilize a larger popular $\mu$-batch, which can be fully executed on GPU, hiding parameter gathering and communication latency for non-popular $\mu$-batches.

\subsection{Comparison: GPU-only Baseline}
\label{subsec:gpu_only_comparison}

\begin{figure}[h!]
	\centering
	\includegraphics[width=0.95\columnwidth]{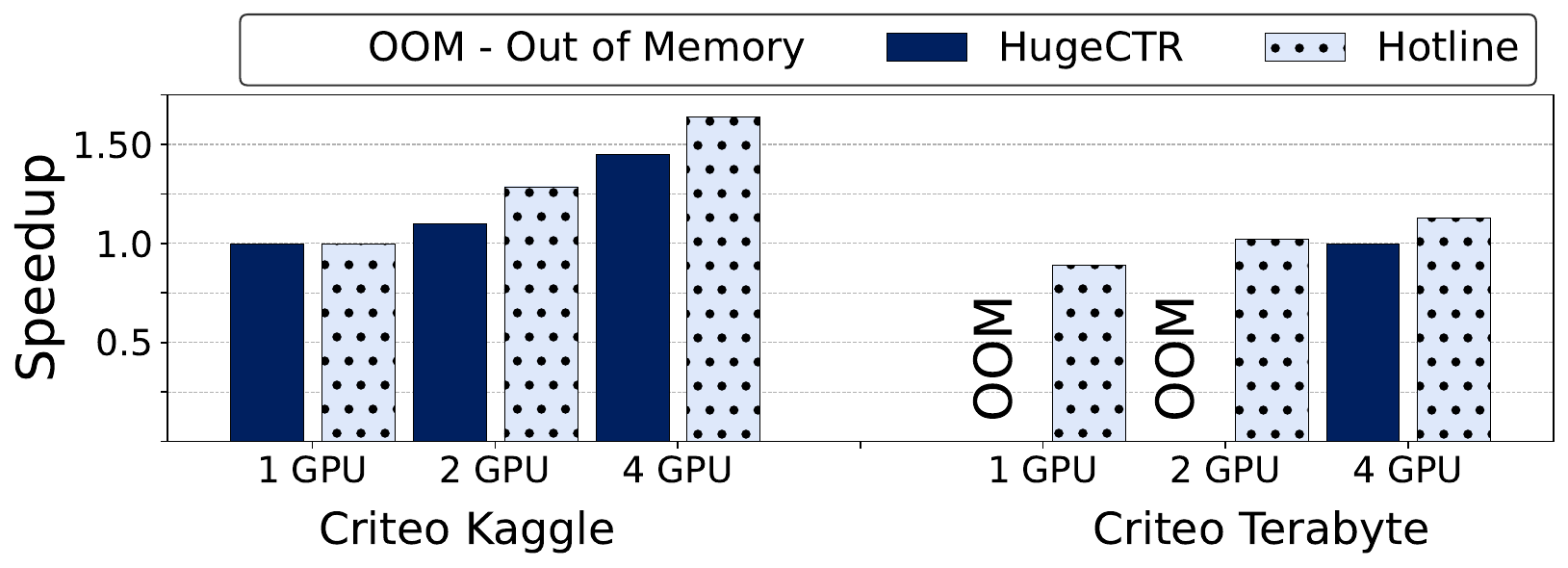}
	\caption{The speedup of \hotline compared to HugeCTR. \hotline eliminates {\fontfamily{qcr}\selectfont{all-to-all}} communication.}
	\label{fig:hugectr_perf}
\end{figure}

We compare \hotline against Nvidia's GPU-only baseline, HugeCTR~\cite{hugectr}. HugeCTR scales the number of GPUs to fit the entire model using model-parallel training for embeddings and data-parallel training for the neural network. Figure~\ref{fig:hugectr_perf} compares Criteo Kaggle and Criteo Terabyte datasets.

HugeCTR can train small models like Criteo Kaggle on a single GPU, so its results are normalized to 1-GPU HugeCTR. However, for large models like Criteo Terabyte, HugeCTR throws an Out of Memory (OOM) error and cannot fit the model within 1 or 2 GPUs, so its results are normalized to 4 GPUs. \hotline eliminates {\fontfamily{qcr}\selectfont{all-to-all}} communication and achieves a speedup of \avgperfimprfourgpuhugectr. 

It is unfair to compare \hotline, a hybrid training scheme, to a GPU-only training scheme. \hotline can train even large datasets such as Terabyte with a single GPU. These datasets would otherwise be unable to be trained on a single GPU. The GPU-only mode needs at least four GPUs for such datasets.

\subsection{Comparison: CPU-based Design}
\begin{figure}[h!]
	\centering
	\includegraphics[width=0.95\columnwidth]{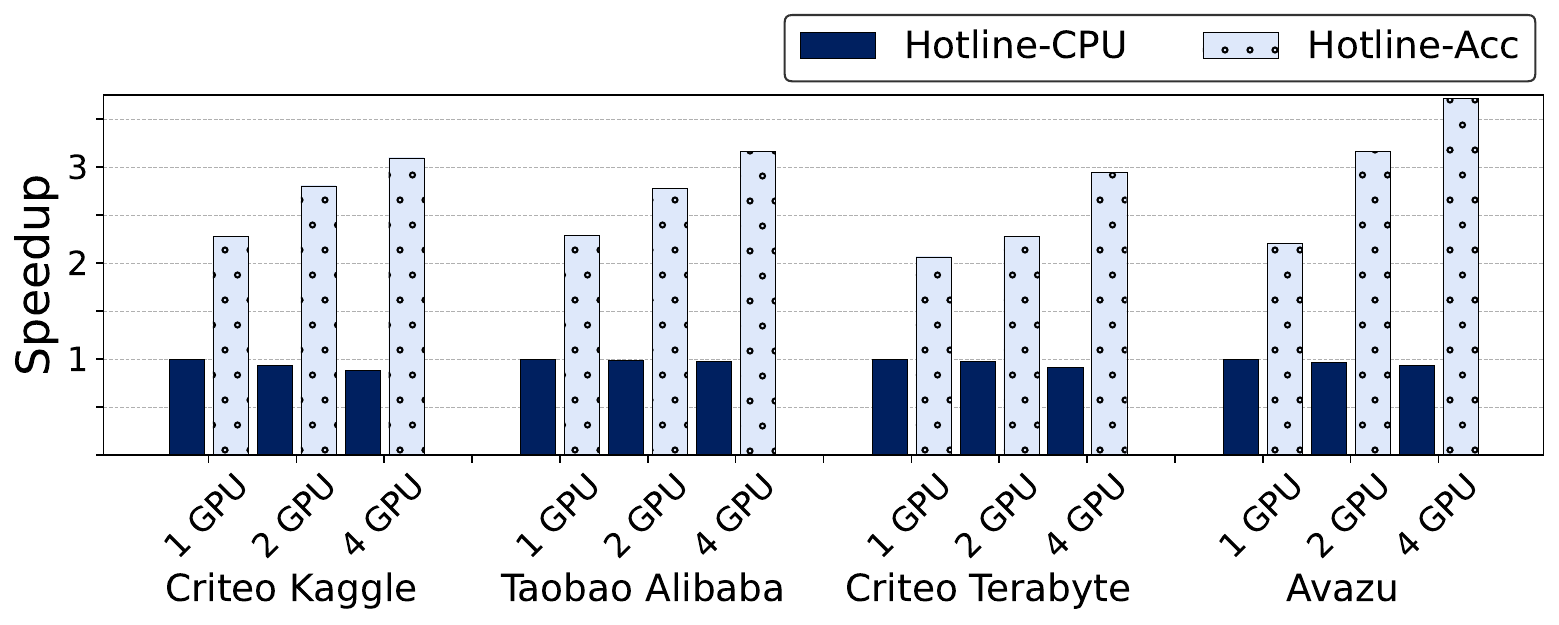}
	\caption{\hotline speedup normalized to CPU-based \hotline implementation. \hotline provides up to 3.5$\times$ higher speedup.}
	\label{fig:hotline_cpu_perf}
\end{figure}

Figure~\ref{fig:hotline_cpu_perf} compares \hotline to a multi-process CPU-based segregator and scheduler. Using the CPU for mini-batch segregation and working parameter gathering results in GPU stalls as the CPU cannot hide the latency behind popular $\mu$-batch execution. \hotline outperforms this alternative approach, providing significant performance benefits.

\subsection{Comparison: Lookahead-Based Software Baselines}

Software-oriented approaches~\cite{scratchpipe, cDLRM, bagpipe} have explored utilizing skewed embedding access patterns by prefetching future mini-batch embeddings into a GPU-based cache. However, such lookahead-based approaches introduce complexities related to data hazards, cache eviction, and model accuracy. For example, cDLRM~\cite{cDLRM} utilizes stale embeddings to mitigate data hazards at the expense of reduced model accuracy.

\begin{figure}[h!]
	\centering
	\includegraphics[width=0.95\columnwidth]{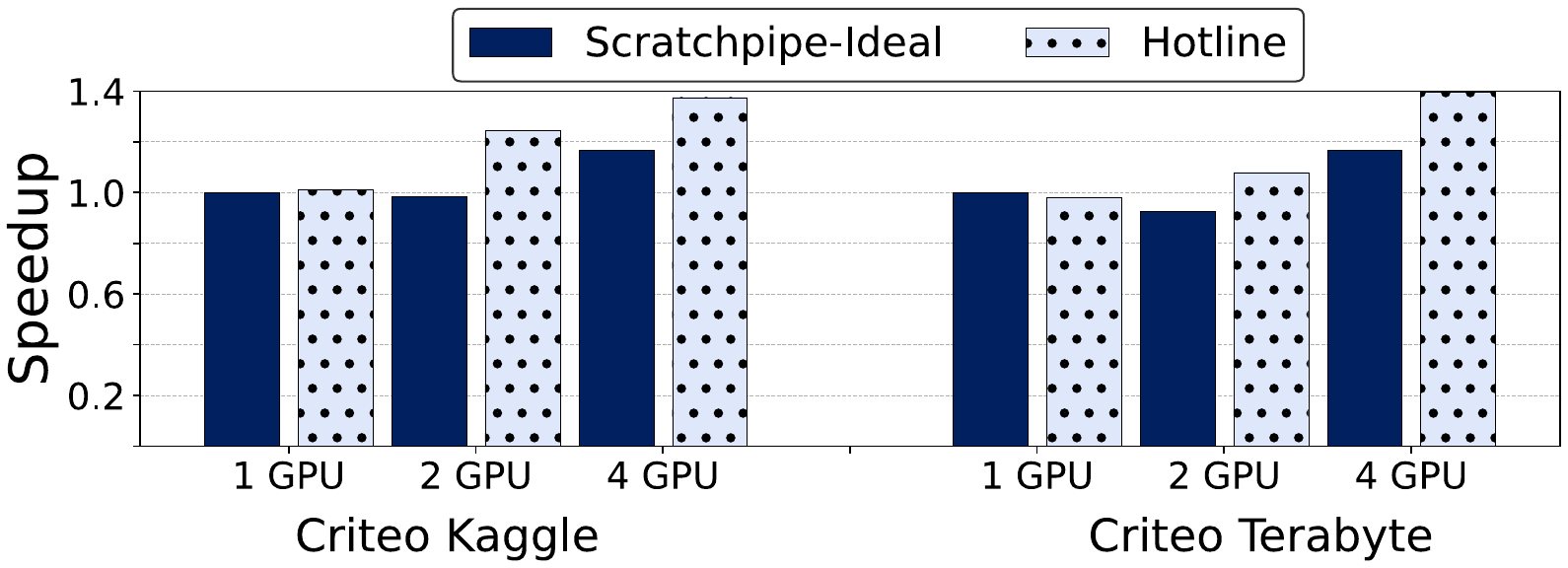}
	\caption{Speedup of \hotline compared to ScratchPipe-Ideal.}
	\label{fig:scratchpipe_perf}
\end{figure}

Figure~\ref{fig:scratchpipe_perf} compares \hotline to ScratchPipe~\cite{scratchpipe}. ScratchPipe~\cite{scratchpipe} is not open-source. Thus, the exact implementation is unknown. Due to this, we re-implemented ScratchPipe with optimistic assumptions and relaxed the stringent read-after-write (RAW) dependencies for model updates. We anticipate a more substantial speedup for Hotline if ScratchPipe adheres strictly to RAW dependencies.
ScratchPipe-Ideal represents an ideal implementation of ScratchPipe~\cite{scratchpipe} with relaxed read-after-write (RAW) dependencies. It performs similarly to \hotline for a single GPU. However, as the number of GPUs increases, ScratchPipe-Ideal encounters scalability issues due to {\fontfamily{qcr}\selectfont{all-to-all}} communication. In contrast, \hotline achieves an average speedup of \avgperfimprfourgpuscratchpipe for 4 GPUs.

\subsection{Sensitivity Studies}
\subsubsection{Varying Popular/Non-Popular $\mu$-batch Ratio}
\label{subsubsec:varying_ratio}
\begin{figure}[h!]
	\centering
	\includegraphics[width=0.95\columnwidth]{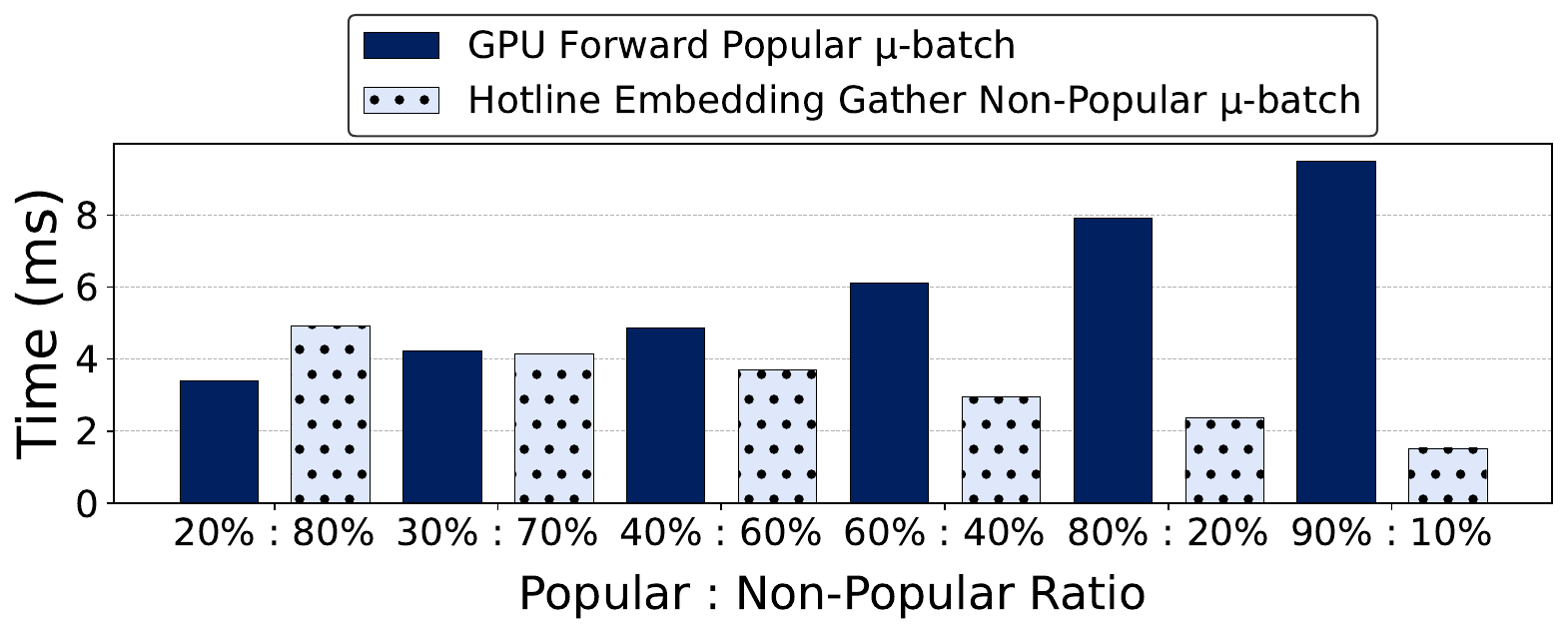}
	\caption{Effect of varying the ratio of popular to non-popular $\mu$-batches. \hotline effectively conceals embedding gather latency even with a \texttt{3:7} popular to non-popular $\mu$-batch ratio.}
	\label{fig:ratio_popular}
\end{figure}

We explored various popular to non-popular $\mu$-batch ratios using a synthetic dataset. Real-world datasets consistently exhibited an average ratio of \texttt{3:1} for popular to non-popular $\mu$-batches, each with 512 MB of \highaccess embeddings. Figure~\ref{fig:ratio_popular} illustrates \hotline's ability to effectively conceal embedding gather latency even with a \texttt{3:7} popular to non-popular $\mu$-batch ratio. Such low ratios are rare in real-world datasets, typically following a Zipfian distribution~\cite{fae,mixedde}.

\subsubsection{Varying Mini-batch Size}
\begin{figure}[h!]
	\centering
	\includegraphics[width=\columnwidth]{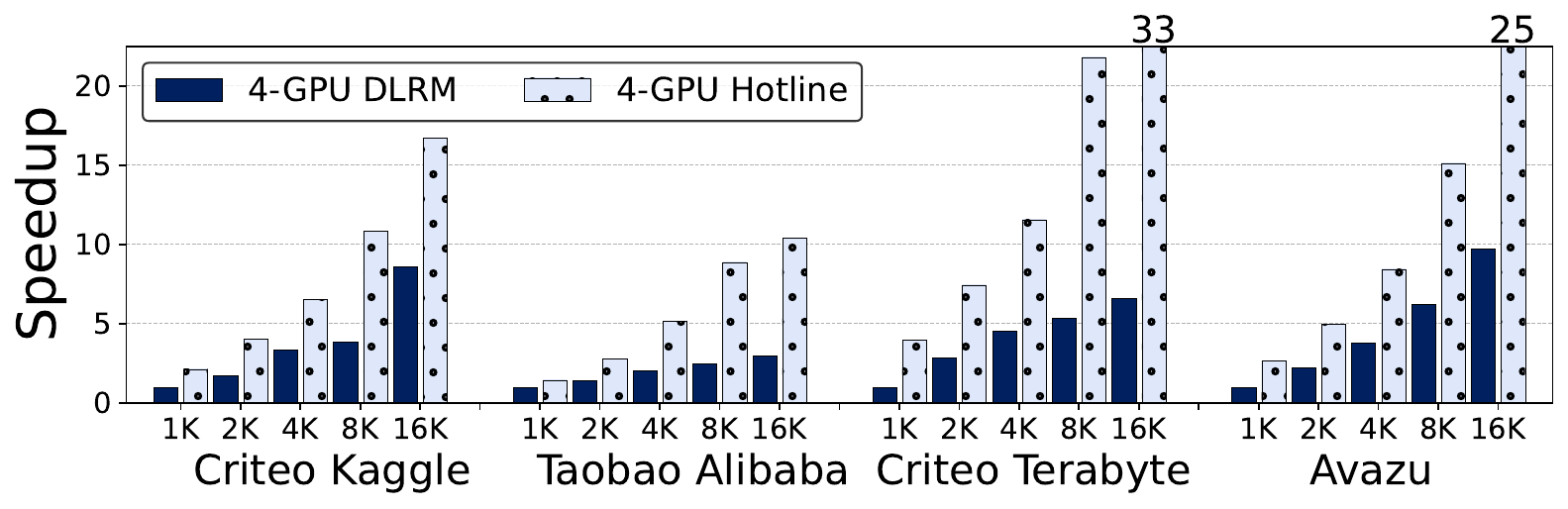}
	\caption{\hotline speedup with varying mini-batch sizes. The benefits of \hotline increase with larger mini-batch sizes.}
	\label{fig:minibatch_perf}
\end{figure}

\hotline benefits increase with larger mini-batch sizes, as shown in Figure~\ref{fig:minibatch_perf}. The scheduler issues fewer input-dispatch commands, and larger mini-batches provide more parallelism opportunities for GPUs. 

\subsubsection{Varying EAL Size}
\begin{figure}[h!]
	\centering	\includegraphics[width=0.9\columnwidth]{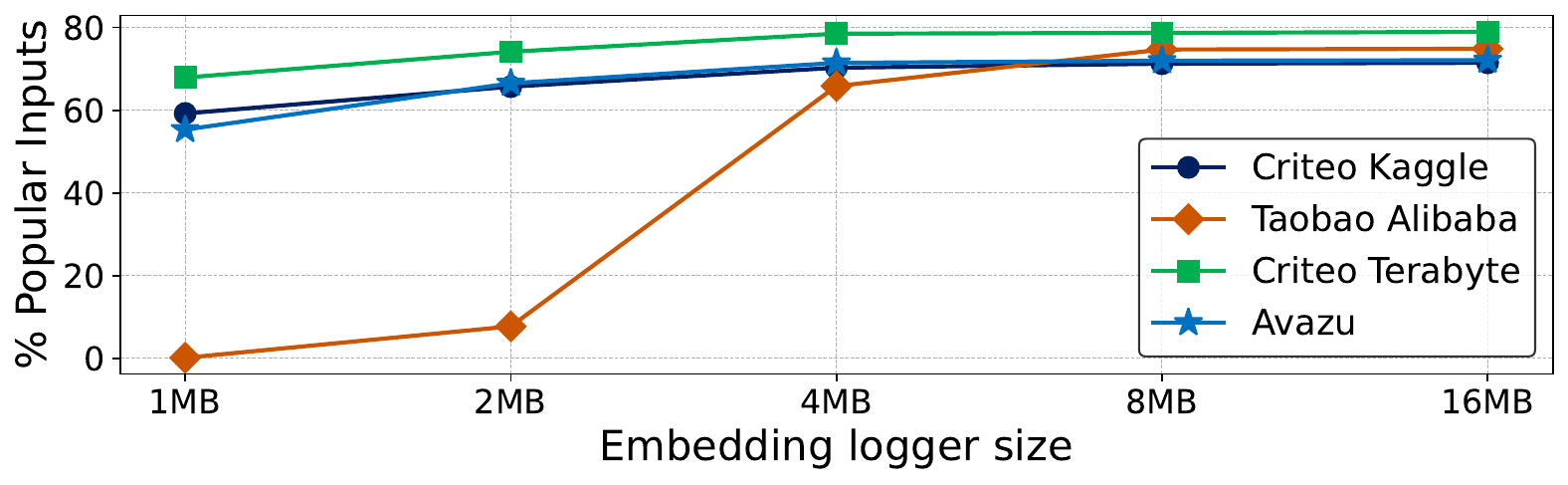}
	\caption{EAL design space exploration shows a 4MB SRAM sufficiently captures the frequently used embedding indices.}
	\label{fig:eal_size}
\end{figure}

Figure~\ref{fig:eal_size} shows popular inputs captured with varying the EAL size. For highly skewed datasets like Criteo and Avazu, a 2MB logger is sufficient to capture \highaccess indices. However, EAL sizes above 4MB offer diminishing returns for the less skewed Taobao dataset.

\subsubsection{Varying Model Size}
\label{subsubsec:syn_models}
We generated synthetic models and datasets with multi-hot encoded inputs to understand the efficacy of \hotline to model size increase. Multi-hot encoded lookups influence the frequency of popular $\mu$-batches. However, due to the heavy-tailed distribution of accesses, over 75\% of these inputs are popular. As shown in Section~\ref{subsubsec:varying_ratio}, this high proportion of popular inputs adequately conceals the parameter gathering latency for non-popular $\mu$-batches.

\begin{figure}[h!]
	\centering
	\includegraphics[width=0.95\columnwidth]{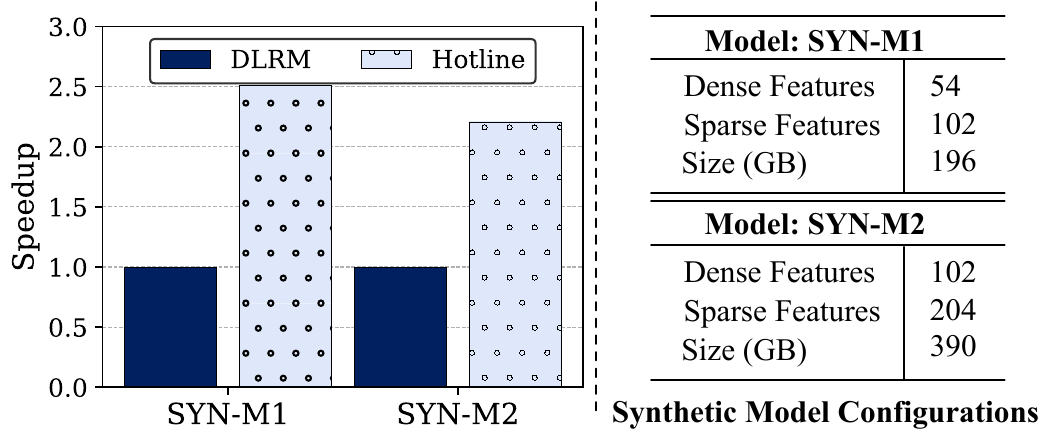}
	\caption{Performance of \hotline versus Intel-Optimized DLRM across synthetic models for a 4-GPU system. The benefits of \hotline are sustained even for larger models.}
	\label{fig:syn_perf}
\end{figure}

Figure~\ref{fig:syn_perf} shows the performance of \hotline across two synthetic models and datasets. Our experiments show that the benefits of \hotline are sustained even for larger models. As the model size increases, the sparse features increase from 102 to 204, and the performance gains decrease from 2.5x to 2.2x. This decrease can be attributed to the \hotline accelerator's parallel lookup engine size remaining the same at 64. With more sparse features, the \hotline accelerator requires more cycles to segregate the input mini-batch, given the fixed size of the parallel lookup engine.

\subsection{Comparison: Performance/Watt, Area, and Power}

\begin{figure}[h!]
	\centering
	\includegraphics[width=0.95\columnwidth]{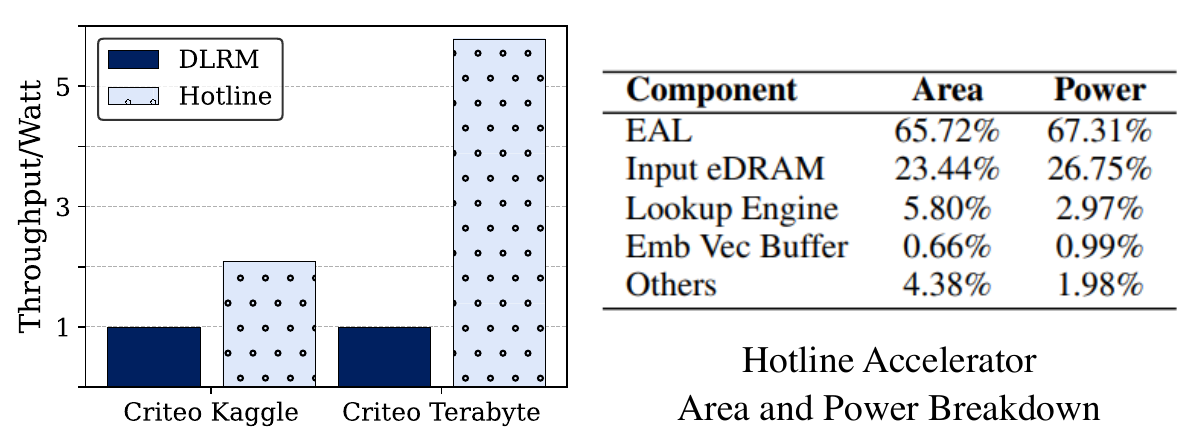}
	\caption{Throughput/Watt comparison with Area and Power breakdown of the \hotline accelerator.}
	\label{fig:perf_w_breakdown}
\end{figure}

Figure~\ref{fig:perf_w_breakdown} shows the Throughput/Watt improvement and area/power consumption of \hotline components. The EAL consumes the most power and area due to its SRAM structure. Despite the 7.01 $mm^2$ area overhead and extra power consumption, \hotline's performance benefits outweigh the power overheads, providing \avgperfW performance/Watt improvement.

\subsection{Multi Node Distributed Training}
\label{subsec:multinode_dist_training}

In the multi-node setup, we evaluated large synthetic models (SYN-M1 and SYN-M2) described in Section~\ref{subsubsec:syn_models}. We compared \hotline to HugeCTR across configurations of 1, 2, and 4 nodes, with mini-batches of 4k, 8k, and 16k. SYN-M1 (196GB) fits only in a 4-node setup, while SYN-M2 (390GB) exceeds the capacity of 4 nodes (16 NVIDIA V100 GPUs).

\begin{figure}[t!]
	\centering
 \includegraphics[width=0.95\columnwidth]{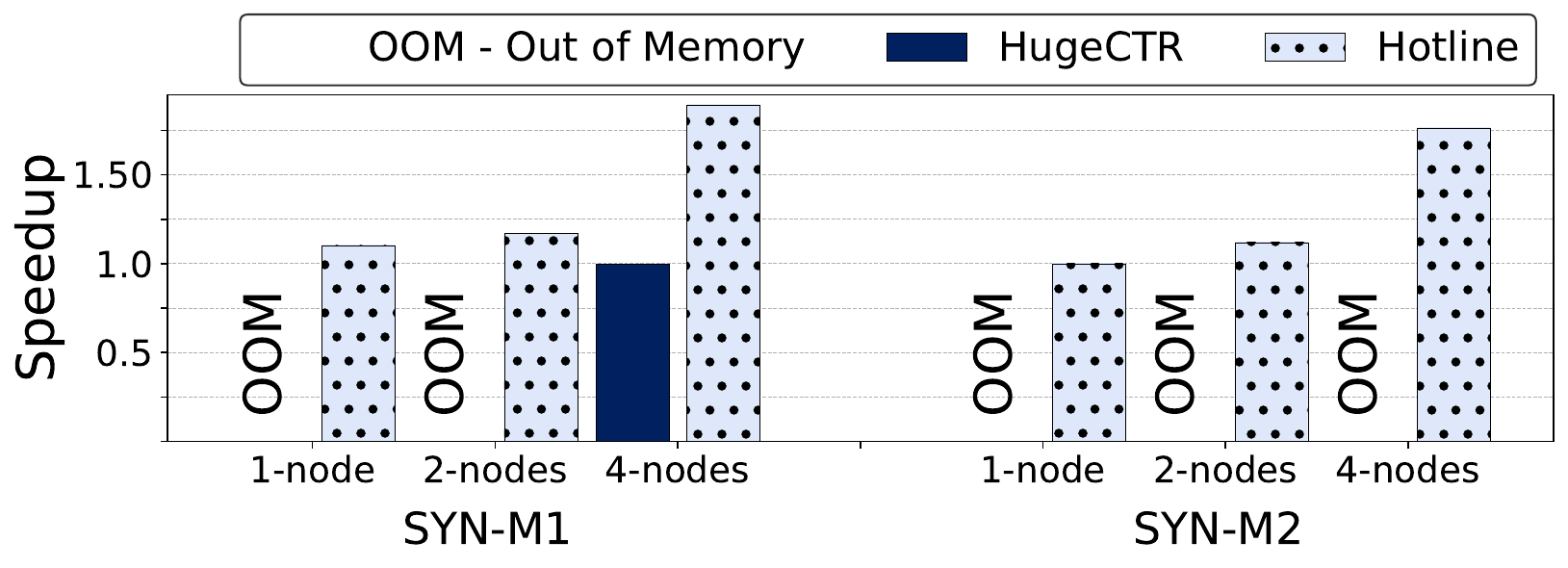}
	\caption{\hotline scalability for multi-node setting with large synthetic models. The speedup is benchmarked with 4-node HugeCTR for SYN-M1 and 1-node \hotline for SYN-M2.}
	\label{fig:multinode_gpu_perf}
\end{figure}

Figure~\ref{fig:multinode_gpu_perf} demonstrates \hotline achieving a 1.89$\times$ speedup on 4 nodes by eliminating {\fontfamily{qcr}\selectfont{all-to-all}} communication, which constitutes over 50\% of total training time. Scaling from one to two nodes encounters challenges due to {\fontfamily{qcr}\selectfont{all-reduce}} synchronization over InfiniBand and CPU-mediated parameter gathering from other nodes. Nevertheless, \hotline's efficiency enables training large models on a single GPU without increasing the GPU count.

\section{Related Work}
\label{sec:related}

\niparagraph{Recommendation models and designs:} Prior work has primarily focused on optimizing the inference phase of recommendation models, as shown in~\cite{centaur, DeepRecSys, nvidiareco, tensordimm, facebookreco:hpca, grace, evstore, mprec}. However, some solutions have also been proposed for training optimizations and acceleration, such as~\cite{mixedde, compemd, recnmp, cosmic:micro}. These solutions, however, do not maximize throughput by effectively utilizing memory and bandwidth in a distributed GPU system. Recently, NEO~\cite{neo} was introduced, which leverages 4-D parallelism for recommendation model training. While NEO can further benefit from embedding placement and patterns in training data, it is orthogonal to \hotline.

\niparagraph{Embedding parameter placement:} Prior methods~\cite{recshard, fae} rely on offline profiling and static embedding placement based on training data skew. In contrast, \hotline dynamically adapts to changing data patterns without such overheads. Recent works~\cite{zhao2020distributed, acun2020understanding} explore alternative embedding table placements but lack preprocessing to reduce communication overheads. Bandana~\cite{bandana} suggests storing embedding tables in non-volatile memory with DRAM caching. Other approaches~\cite{mixedde, recnmp, compemd, softwarehardware} accelerate near-memory processing but lack support for distributed training with GPUs. Prior work~\cite{dreamshard, autoshard} performs embedding placement across GPU-device for GPU-only training, while \hotline targets a two-tier memory hierarchy for hybrid training.

\niparagraph{Mitigating memory intensive training:} Prior work has focused on optimizing the model using mixed-precision training or eliminating rare categorical variables to reduce embedding table size~\cite{nvopt, mixedprecision}. However, changing the data representation or embedding tables requires accuracy re-validation. Compression and sparsity have also been used to reduce model memory footprint~\cite{compressreco, gist, compressreco2, sparsemat, elrec}. In contrast, \hotline performs full-precision training without the overheads of compression/decompression and sparse operations. It only leverages access skew and is independent of these techniques.

\niparagraph{Embedding Representation:} Previous research has explored various methods to represent categorical features within limited memory, aiming to accommodate multiple feature values with a restricted number of embeddings. The hashing trick~\cite{hashing} applies a simple hash function to constrain feature embeddings. Compositional Embeddings~\cite{compemd} leverages complementary partitions of categorical features, utilizing multiple smaller embedding tables and combining embeddings from each table. ROBE~\cite{robe} accesses contiguous blocks in shared memory for enhanced memory access. Unified Embeddings~\cite{unified} consolidates all categorical features within a single embedding table, allowing for collisions of feature values within and across features. DHE~\cite{dhe} employs an orthogonal approach, representing feature values using MLPs and encoders instead of embeddings. Hotline can be applied atop any embedding representation technique.

\niparagraph{Embedding Prefetching:} Previous studies~\cite{cDLRM, scratchpipe, bagpipe} have investigated prefetching embeddings into a GPU-based cache for the next mini-batch of training. However, this prefetching-based approach introduces complexities such as data hazards, complex cache eviction policies, and asynchronous training with limited scalability. In contrast, Hotline avoids these complexities through pipeline scheduling within a minibatch.

\niparagraph{Machine learning accelerators:} There are proposals for accelerators designed to execute the compute portion of deep learning models~\cite{tpu, dnnweaver:micro:2016, minerva, eyeriss, brainwave}, including some for collaborative filtering-based recommender models~\cite{tabla:hpca, cosmic:micro, dana}. However, \hotline does not aim to design a specialized architecture for optimizing the computing of deep learning-based recommender models. \hotline accelerator can pipeline and potentially enhance these existing accelerators.

\section{Conclusions}
\label{sec:conclusion}
This paper proposes \hotline, a heterogeneous acceleration pipeline to address memory and bandwidth constraints in recommendation models. \hotline leverages the insight that only a few embedding table entries are popular and frequently accessed. It sends inputs directly to the GPUs, which use \highaccess embeddings. It retrieves the required embeddings for the remaining inputs while the GPUs process popular inputs. \hotline uses a novel accelerator to dynamically segregate and dispatch the data, hiding the data transfer latency behind the GPU execution of popular inputs. Our experiments on real-world datasets and models show that \hotline reduces average training time by \avgperfimprfourgpudlrm compared to Intel-optimized CPU-GPU DLRM baseline.
\balance
\section*{Acknowledgements}

This project is part of the \textit{STAR Lab} at The University of British Columbia (UBC). We thank the entire Advanced Research Computing Center team at UBC~\cite{sockeye}. We also thank the anonymous reviewers from ISCA 2024 for their invaluable feedback. Muhammad Adnan's Ph.D. is supported by Intel Transformation Server Architecture (TSA) and the Natural Sciences and Engineering Research Council of Canada (NSERC) [funding reference number RGPIN-2019-05059] Grants. The views and conclusions contained herein are those of the authors. They should not be interpreted as representing the official policies or endorsements, expressed or implied, of NSERC, the Canadian Government, Georgia Tech, or The University of British Columbia.

\bibliographystyle{unsrtnat}
\bibliography{refs}

\end{document}